\documentclass[sort&compress]{elsarticle}

\usepackage{fullpage}
\usepackage{amssymb}
\usepackage{array}  
\usepackage{pifont}
\usepackage{bbding}
\usepackage{graphicx}
\usepackage{subfig}
\usepackage{enumerate}
\usepackage{amsmath,mathtools}
\usepackage{amsfonts,bm}
\usepackage{todonotes}
\usepackage[commentmarkup=footnote]{changes}
\usepackage{stmaryrd}
\usepackage{siunitx}
\usepackage{caption}
\usepackage{array}
\usepackage{booktabs}
\usepackage{nomencl}
\usepackage{lineno}
\usepackage{url}
\usepackage{natbib}
\usepackage[strings]{underscore}
\newcommand{\bbR}{\mathbb {R}}
\newcommand{\bx}{\bm{x}}

\newcommand{\bu}{\mathbf{u}}

\newcommand{\bc}{\bm{c}}

\newcommand{\cE}{\mathcal{E}}
\newcommand{\cO}{\mathcal{R}}

\newcommand{\cQ}{\mathcal{Q}}
\newcommand{\cG}{\mathcal{G}}
\newcommand{\cX}{\mathcal{X}}

\newcommand{\cD}{\mathcal{D}}

\newcommand{\bigcheck}{\text{\large\checkmark}}

\newcolumntype{C}[1]{>{\centering\arraybackslash}p{#1}}

\renewcommand\nomgroup[1]{%
  \item[\itshape
  \ifstrequal{#1}{A}{Symbols}{%
  \ifstrequal{#1}{B}{Roman Letters}{%
  \ifstrequal{#1}{C}{Greek Letters}{%
  \ifstrequal{#1}{D}{Abbreviations}{}}}}%
]}

\usepackage{xpatch}
\xpatchcmd{\thenomenclature}{\section*{\nomname}
}{}{\typeout{Success}}{\typeout{Failure}}
\makenomenclature
\RequirePackage{ifthen}
\printnomenclature[0.8cm]

\usepackage[commentmarkup=footnote]{changes}
\definechangesauthor[name={Reviewer 1}, color = red]{R1}
\definechangesauthor[name={Reviewer 2}, color = brown]{R2}
\definechangesauthor[name={Author}, color = orange]{Author}
\usepackage{xcolor}

\journal{Journal of Computational Physics}

\graphicspath{{./figs/}}

\begin{document}

\begin{frontmatter}

\title{Neural operator-based super-fidelity: A warm-start approach for accelerating steady-state simulations\tnoteref{titlefn}}

\author[vt]{Xu-Hui Zhou\corref{cor1}}%
\ead{xuhuizhou@vt.edu}
\author[fi]{Jiequn Han}%
\author[vt]{Muhammad I. Zafar}%
\author[oa]{Eric M. Wolf}
\author[af]{\\Christopher R. Schrock}%
\author[vt]{Christopher J. Roy}%
\author[st]{Heng Xiao}%

\cortext[cor1]{Corresponding author.}
\tnotetext[titlefn]{Distribution Statement A: Approved for Public Release; Distribution is Unlimited. PA\# AFRL-2024-0850.}

\affiliation[vt]{organization={Kevin T. Crofton Department of Aerospace and Ocean Engineering, Virginia Tech},
            city={Blacksburg},
            postcode={24060}, 
            state={VA},
            country={USA}}
\affiliation[fi]{organization={Center for Computational Mathematics, Flatiron Institute},
            city={New York},
            postcode={10010}, 
            state={NY},
            country={USA}}
\affiliation[oa]{organization={Ohio Aerospace Institute, Wright-Patterson Air Force Base},
            city={Dayton},
            postcode={45433}, 
            state={OH},
            country={USA}}
\affiliation[af]{organization={Air Force Research Laboratory, Wright-Patterson Air Force Base},
            city={Dayton},
            postcode={45433}, 
            state={OH},
            country={USA}}
\affiliation[st]{organization={Stuttgart Center for Simulation Science, University of Stuttgart},
            city={Stuttgart},
            postcode={70569}, 
            state={BW},
            country={Germany}}

\begin{abstract}

Neural networks have recently emerged as powerful tools for accelerated solving of partial differential equations (PDEs) in both academic and industrial settings. However, their use as standalone surrogate models raises concerns about reliability, as solution accuracy heavily depends on data quality, volume, and training algorithms. This concern is particularly pronounced in tasks that prioritize computational precision and deterministic outcomes. In response, this study introduces ``super-fidelity'', a method that employs neural networks for initial warm-starts, significantly speeding up the solution of steady-state PDEs without compromising on accuracy. Drawing from super-resolution in computer vision, super-fidelity maps solutions from low-fidelity computational models to high-fidelity ones using a vector-cloud neural network with equivariance (VCNN-e)---a neural operator that preserves physical symmetries and adapts to different spatial discretizations. 
We evaluated the proposed method across scenarios with varying degrees of nonlinearity, including (1) two-dimensional laminar flows around elliptical cylinders at low Reynolds numbers, exhibiting monotonic convergence, (2) two-dimensional turbulent flows over airfoils at high Reynolds numbers, characterized by oscillatory convergence, and (3) practical three-dimensional turbulent flows over a wing. 
The results demonstrate that our neural operator-based initialization can accelerate convergence by at least a factor of two while maintaining the same level of accuracy, outperforming traditional initialization methods using uniform fields or potential flows. The approach's robustness and scalability are confirmed across different linear equation solvers and multi-process computing configurations. Additional investigations highlight its reduced dependence on high quality of training data, and real time savings across multiple simulations, even when including the neural-network model preparation time. Our study presents a promising strategy for accelerated solving of steady-state PDEs using neural operators, ensuring high accuracy in applications where precision is of utmost importance.
\end{abstract}

\begin{keyword}
neural network/operator \sep partial differential equations  \sep acceleration \sep super-fidelity \sep warm-start

\end{keyword}

\end{frontmatter}

\section{Introduction}
\label{sec:intro}
Partial differential equations (PDEs) serve as a fundamental and ubiquitous tool to describe how various properties change continuously in space and time. Their significance extends to numerous fields, including physics, engineering, finance, and beyond. While the development of numerical methods has enabled efficient solutions for certain types of PDEs, there remains a subset of PDEs for which traditional techniques are computationally intensive, thus limiting their applicability in tackling complex problems.
Recent years have witnessed a promising new paradigm in scientific computing, as neural networks have emerged as promising tools for PDE solutions. Neural networks, with the capacity to approximate complex functions, can deliver near-ground-truth predictions quickly once well trained. The latest development of neural operators has further extended this capability to infinite-dimensional function spaces, enabling trained models to adapt to inputs with arbitrary resolutions/discretizations; see, e.g., \cite{han2019uniformly,lu2021learning,li2021fourier,kovachki2023neural,han2023equivariant,chen2023operator,huang2024operator,zhou2024bi}. This paradigm shift in computational science has brought us more insights and new possibilities for solving PDEs.

However, relying on neural network predictions as the ultimate solution to PDEs often triggers a trust crisis, especially for tasks that prioritize computational precision and outcome determinism. This trust issue exists in two prevalent neural network-based approaches for PDE solving: data-driven methods, such as supervised learning of solution operators, and physics-driven methods, such as physics-informed neural networks (PINNs). In the case of data-driven methods, training an accurate and reliable network model is highly challenging, which significantly depends on data volume and quality, as well as training algorithms. While studies have shown that well-trained models can produce results with desired accuracy in certain cases~\cite{shukla2023deep,chen2023towards}, the inherent black-box nature of neural networks and the absence of physical constraints (e.g., conservation laws) during training render these predictions impractical for researchers who rely on first principles in their work. On the other hand, PINNs aim to solve PDEs by training neural networks to satisfy physical constraints~\cite{raissi2019physics}. However, comparative studies on PINNs and other machine learning-based solvers versus traditional numerical methods indicate that machine learning solvers do not exhibit a clear advantage either in terms of accuracy or computational efficiency for various PDEs in low-dimensional cases~\cite{grossmann2023can, mcgreivy2024weak}. Therefore, for scientific computing tasks that require solution precision, traditional solvers should continue to occupy a central role. At this stage, it is essential to mitigate the risk of misusing neural network predictions and ensure the accuracy of the results.

In this paper, we propose an approach that employs neural network predictions as initial conditions to accelerate the solution of steady-state PDEs. We have chosen steady-state fluid flow problems to demonstrate the efficacy of our approach, considering their industrial relevance and the inherent complexity associated with computational models based on the Navier--Stokes (N--S) equations.
Specifically, our proposed strategy involves the development of an approximate mapping from the solution of a low-fidelity model (i.e., Laplace's equation for potential flow) to the solution of a high-fidelity model (i.e., the target PDEs). This mapping, which is straightforward and fast to evaluate, provides a solution that is much closer to the final solution than the low-fidelity model. By utilizing the output of this mapping as an improved initial guess, we can effectively warm-start the iterative solution process, further reducing the total computational time. We have referred to this task as ``super-fidelity'', drawing inspiration from the super-resolution concept in image processing, which enhances a low-resolution image to achieve higher resolution~\cite{dong2015image}. The objective of super-fidelity is to bridge the gap between low and high-fidelity models, thereby enhancing the efficiency and stability of steady-state simulations in complex flows. In this context, our approach conceptually aligns with previous ``solver-in-the-loop'' efforts~\cite{kochkov2021machine,um2020solver}, which focus on accelerating time-dependent PDE solutions by bridging the gap between coarse-grid and fine-grid solutions. For example, study~\cite{kochkov2021machine} employs machine learning to deliver pointwise accurate solutions on unresolved grids, allowing computations on coarse grids while preserving accuracy and reducing computational costs. Similarly, study~\cite{um2020solver} integrates a neural network-based corrector into the iterative process of PDE solvers, dynamically refining coarse-grid solutions to achieve fine-grid accuracy with lower computational expense. Despite these parallels, our approach targets a different application scenario, focusing specifically on accelerating steady-state simulations via improved initial conditions. This distinction is particular relevant for industrial applications, where steady-state solutions are critical for performance evaluation and operational efficiency.

\begin{figure}[!htb]
\centering
\includegraphics[width=0.99\textwidth]{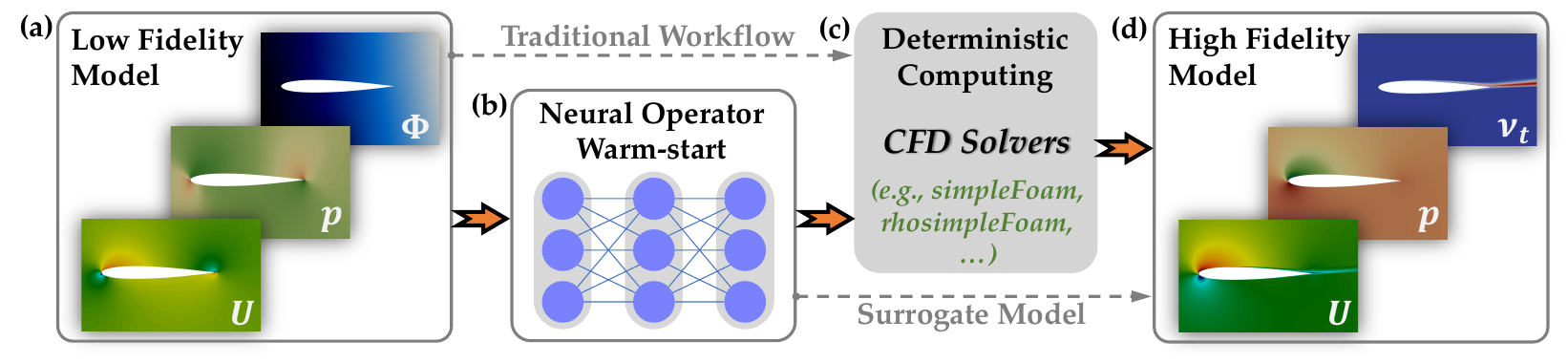}
  \caption{
  Schematic workflow of accelerating steady-state fluid flow simulations with the neural operator warm-start approach. The neural operator is trained to map solutions from a low-fidelity model to a high-fidelity model and then provides improved initial conditions for CFD simulations. This approach transcends traditional workflows by delivering notably enhanced initial conditions to reduce the computational time. It also bolsters the reliability of employing neural networks for scientific problems by performing the deterministic computing afterwards, in contrast to using neural networks purely as surrogate models.
  }
  \label{fig:workflow}
\end{figure}

In the context of warm-starting existing iterative solvers for high-fidelity models, our focus lies in constructing a mapping from low-fidelity solutions to high-fidelity ones---the core component of our super-fidelity approach. This mapping, mathematically, is also referred to as an operator between two function spaces. The field of operator learning, which has received significant research interest in recent years, offers such a way forward. This approach aims to develop a data-driven approximation of mappings/operators between function spaces to provide efficient evaluations of the target output after an initial training phase on available data. In particular, neural networks with specialized architectures, also known as neural operators, have shown considerable promise in approximating such mappings.
In this work, we utilize our recently developed neural operator, the vector-cloud neural network with equivariance (VCNN-e)~\cite{zhou2022frame,han2023equivariant}, to establish the desired mapping in super-fidelity. 
While other neural operators, such as the Fourier neural operator~\cite{li2021fourier}, are also capable of handling this task, we have selected VCNN-e due to its ability to strictly guarantee physical symmetry, allowing it to be applied in different coordinate systems seamlessly.
Furthermore, it considers the nonlocal property of flow simulations, enabling learning to be done with just a few solutions by decomposing the whole solution field into many patches for efficient learning. The entire workflow of our super-fidelity approach includes all four panels shown in Fig.~\ref{fig:workflow}. The second panel, referred to as the warm-start, involves the training of a super-fidelity model based on neural operators. Our numerical experiments on steady-state fluid flow simulations with varying degrees of nonlinearity illustrate the effectiveness of the super-fidelity approach. The warm-start models consistently achieve acceleration ratios of no less than twofold without any compromise on accuracy. For practitioners in computational fluid dynamics (CFD) who typically employ uniform fields or solutions from lower-fidelity models as initial conditions (see ``traditional workflow'' in Fig.~\ref{fig:workflow}, including the (a), (c), and (d) panels), trained neural networks can deliver better initial conditions, thereby speeding up the solving process. This is particularly beneficial to problems involving multiple simulations under varying conditions, a common need in addressing optimal design or control problems in diverse engineering applications~\cite{li2019data}. 
For researchers devoted to accelerating PDE solving through machine learning (see ``surrogate model" in Fig.~\ref{fig:workflow}, including the (a), (b), and (d) panels), integrating traditional solvers as the final touch ensures the accuracy of results.

The remainder of this paper is organized as follows. Section~\ref{sec:method} introduces the computational models used in this study, including the governing equations for both high-fidelity and low-fidelity models, and describes the neural operator architecture. In Section~\ref{sec:result}, we present the neural operator predictions and investigate the resulting acceleration effects for numerical solvers on three test cases. In Section~\ref{sec:discuss}, we introduce an efficient strategy for data generation and explore potential application scenarios to achieve a ``real" acceleration by reducing overall computation time, which includes both data generation and training phases.
We also discuss the robustness and scalability of our method, as well as the limitations of developing reliable neural network surrogates. Section~\ref{sec:conclude} provides a summary of this work and some concluding remarks.

\section{Problem Statement and Methodology}
\label{sec:method}

\subsection{Physical model}
In this study, we assume the high-fidelity computational models for steady-state flow fields adhere to the following conservation law, presented in a general and compact form:
\begin{equation}
\nabla \cdot \mathcal{F}(\bm{U}) - \mathcal{S} = 0,
\label{eq:high-fidelity}
\end{equation}
where $\bm{U}$ is the conserved variable vector and $\mathcal{F}$ and $\mathcal{S}$ denote the flux vector and the source vector, respectively. Taking N--S equations as an example, if we neglect the influence of body forces and radiative heating (i.e., $\mathcal{S} = 0$), the vectors are defined as follows,
\begin{equation}
\bm{U}=\begin{bmatrix} \rho \\ \rho \bm{u} \\ \rho e_{t} \end{bmatrix}, \qquad \mathcal{F}(\bm{U}) = \begin{bmatrix} \rho \bm{u}^\top \\  \rho \bm{u} \otimes \bm{u} + p \bm{I} - \bm{\tau} \\ (\rho e_t + p)\bm{u}^\top - \bm{u}^\top \bm{\tau} + \bm{q}^\top  \end{bmatrix},
\label{eq:NS-Eq}
\end{equation}
where $\rho$, $\bm{u}$, $p$, $\bm{q}$ and $e_t$ denote the fluid's density, velocity, pressure, heat flux, and total energy per unit mass, respectively; $\bm{\tau}$ is the viscous stress tensor and $\bm{I}$ is the identity tensor. The system of equations has to be completed by a thermodynamic equation of state (EOS) in a general form of $p = p(\rho, e)$~\cite{ma2023efficient}, where the internal energy per unit mass $e = e_t - \frac{1}{2}|\bm{u}|^2$. 

Note that Eq.~\eqref{eq:NS-Eq} varies according to specific fluid flow simulation scenarios. For instance, in the case of an incompressible laminar flow, where density is assumed to be constant, the energy equation involving the third components in $\bm{U}$ and $\mathcal{F}$ can often be decoupled from the mass and momentum equations, and can be further neglected if the flow is isothermal. Consequently, the problem simplifies, and only velocity and pressure need to be solved for.

The low-fidelity model considered in this work is based on the potential flow theory, which represents idealized fluid motion. Specifically, the fluid is assumed to be inviscid, while the flow itself is considered irrotational and incompressible. The steady-state potential flow is governed by Laplace's equation~\cite{hirsch2007numerical}
\begin{equation}
\nabla^2 \Phi = 0, \quad \text{with }\ \bm{u} = \nabla \Phi,
\label{eq:low-fidelity}
\end{equation}
where $\Phi$ is the velocity potential. Despite its simplifications and limitations in modeling real fluid behavior, potential flow provides rapid simulations for studying various fluid flow phenomena, typically within seconds.
This makes it an attractive initialization option in steady-state fluid flow simulations, with the aim of improving stability and convergence speed, as seen in simulation software such as Ansys~\cite{ansys2013ansys} and OpenFOAM~\cite{opencfd21openfoam}. In view of the physical information contained within, as well as its cost-effectiveness, it is natural to use potential flow as input for the super-fidelity model. Therefore, the super-fidelity task is to train a neural operator to map the solution of Eq.~\eqref{eq:low-fidelity} to that of Eq.~\eqref{eq:high-fidelity}.

\subsection{Equivarient neural operator}
The neural operator employed in the super-fidelity task is developed based on the VCNN-e~\cite{han2023equivariant} to satisfy all the symmetries and function space requirements for predicting both scalar and vector fields. The neural operator forms a region-to-point mapping, $\mathcal{Q} \mapsto S(\bm{x}_0)$, where $S(\bm{x}_0)$ is the solution of a high-fidelity model at a point of interest $\bm{x}_0$, and $\mathcal{Q}$ indicates the collection of potential flow (low-fidelity model) features, $\{\mathbf{q}_i\}_{i=1}^n$, on $n$ points sampled from the computational domain. The feature vector $\mathbf{q}$ attached to each point is chosen to include the relative coordinate $\bm{x}'=\bm{x}-\bm{x}_0$, potential flow velocity $\bm{u}^L$, and additional six scalar quantities $\bm{c}^L = [\mathsf{u}^L,p^L,\Phi,\theta,\eta,r]$, with the superscript $L$ denoting the low-fidelity model. In scalar quantities, $\mathsf{u}$ and $p$ represent velocity magnitude and pressure; $\theta$ and $\eta$ denote cell volume and wall distance; and $r$ is proximity (inverse of relative distance) to the point of interest.
As such, the feature vector $\mathbf{q}$ is $2 + 2 + 6 = 10$ dimensional for two-dimensional flows, and we define the input matrix as $\mathcal{Q} = [\mathbf{q}_1, \ldots, \mathbf{q}_n]^\top \in \mathbb{R}^{n \times 10}$. Similarly, for three-dimensional flows, the feature vector has $3 + 3 + 6 = 12$ dimensions, and the input matrix is defined as $\mathcal{Q} \in \mathbb{R}^{n \times 12}$.
The output of the neural operator includes the solutions of velocity $\bm{u}^H$ and other scalar quantities $\bm{c}^H$ (e.g., pressure) of a high-fidelity model, i.e., $S(\bm{x}_0) = [\bm{u}^H(\bm{x}_0), \bm{c}^H(\bm{x}_0)]$. Note that all input and output of the neural operator are non-dimensionalized using characteristic scales and normalized.

\begin{figure}[!htb]
\centering
\includegraphics[width=0.99\textwidth]{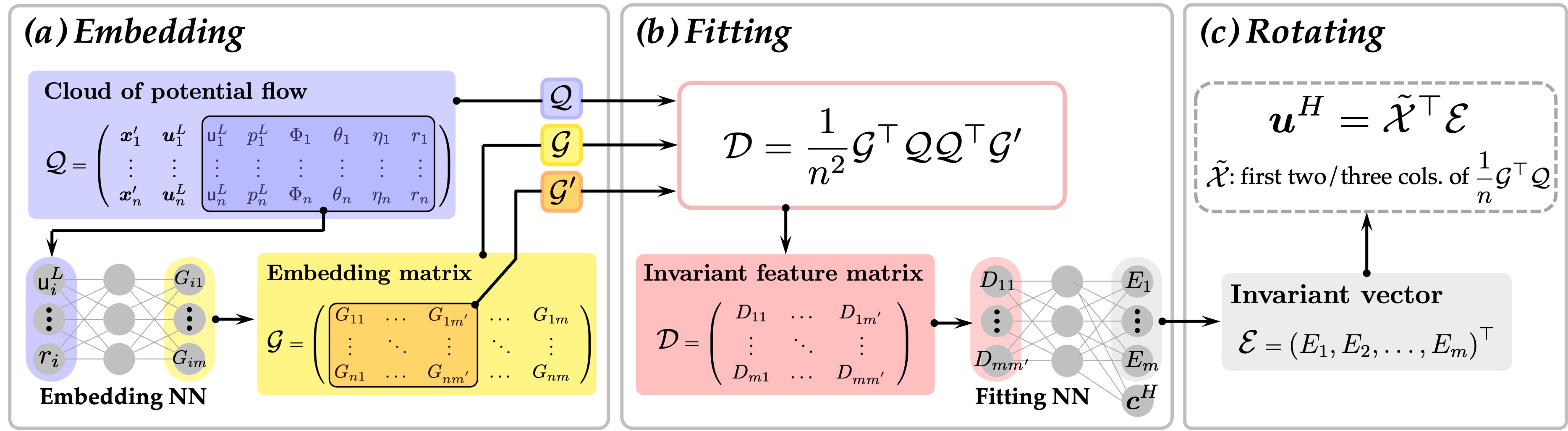}
  \caption{
  Detailed architecture of the neural operator VCNN-e for mapping solution of a low-fidelity model (potential flow) to that of a high-fidelity model:
  (a) embed the frame-independent features $\{\bm{c}_i\}_{i=1}^n$ to form the embedding matrix $\mathcal{G} \in \mathbb{R}^{n \times m}$;
  (b) project the pairwise inner-product matrix $\mathcal{Q} \mathcal{Q}^\top$ to the learned embedding matrix $\mathcal{G}$ and its submatrix $\mathcal{G}^\prime$ to yield an invariant feature matrix $\mathcal{D} \in \mathbb{R}^ {m \times m'}$;
  flatten and feed the feature matrix $\mathcal{D}$ into the fitting network to predict the invariant vector $\cE$ and scalar quantities $\bm{c}^H$; 
  (c) rotate $\cE$ through the embedded coordinates $\tilde{\cX}$ and get the prediction of velocity $\bm{u}^H$.
  The final output consists of $\bm{c}^H$ in (b), which is invariant to the frame rotation, and $\bm{u}^H$ in (c), which is equivariant to the frame rotation, while both of them are invariant to the frame translation and the ordering of points in the cloud.
  }
  \label{fig:NN-arch}
\end{figure}

The neural operator VCNN-e consists of three modules at a high level: (a) an embedding module, (b) a fitting module, and (c) a rotating module, as illustrated in Fig.~\ref{fig:NN-arch}.
The first module (Fig.~\ref{fig:NN-arch}(a)) intends to find an invariant representation, denoted as $\cD$, for the input cloud of potential flow, which has translational, rotational, and permutational invariance. Specifically, the translational invariance is ensured by the use of relative coordinates rather than absolute ones; the rotational invariance is ensured by the inner product between feature vectors via the operation $\cQ \cQ^\top$; and the permutational invariance and resolution adaptivity are guaranteed by the average over all the points via the operations $\frac1n\cG^\top \cQ$ and $\frac1n\cQ^\top \cG'$.
The second module utilizes $\cD$, referred to as the invariant feature matrix, as the input to fit both an invariant vector $\cE$ and predictions of scalar quantities $\mathbf{c}^H$ for the high-fidelity model through a fitting neural network.
Finally, the third module (Fig.~\ref{fig:NN-arch}(c)) incorporates the embedded coordinates $\tilde{\cX} = \frac{1}{n} \cG^\top \cX$ from the first module where $\cX = [\bx'_1, \ldots, \bx'_n]^\top$ and rotates properly to get the velocity prediction through $\bm{u}^H = \tilde{\cX}^\top \cE$. Since $\cX$ rotates in the same way as the original coordinates under rotation, the predicted velocity vector preserves the desired equivariance. Note that the output of the neural operator consists of $\bm{u}^H$ from the rotating module and $\bc^H$ from the fitting module, which together serve as the initial conditions for simulating the high-fidelity model. The neural operator architecture is detailed in~\ref{app:NN-arci}. More details regarding the design of VCNN-e and the proof of symmetries can be found in the literature~\cite{zhou2022frame,zafar2022frame,han2023equivariant}.

\section{Results}
\label{sec:result}

\begin{figure}[!htb]
\centering
\includegraphics[width=0.8\textwidth]{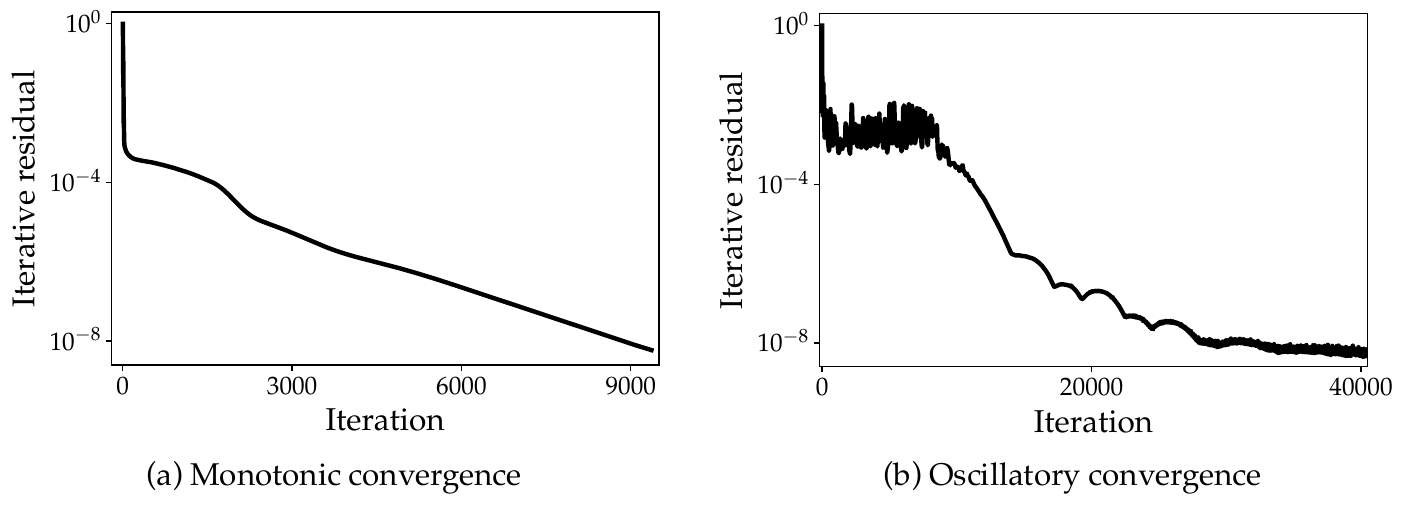}
  \caption{
  Two typical convergence processes in scientific computing: (a) monotonic convergence and (b) oscillatory convergence.
  }
  \label{fig:two-convergence}
\end{figure}

In this section, we present a comprehensive evaluation of our warm-start approach by applying it to three different flow scenarios.
The first scenario examines two-dimensional incompressible laminar flows over parameterized elliptical cylinders, while the second and third scenarios investigate two-dimensional and three-dimensional compressible turbulent flows over airfoils and a wing, respectively, under varying flow conditions.
The convergence behaviors in these flow scenarios generally align with the typical patterns commonly observed in scientific computing~\cite{oberkampf2010verification,phillips2014richardson}, as illustrated in Fig.~\ref{fig:two-convergence}. The first scenario demonstrates weaker nonlinearity with low Reynolds numbers, leading to a monotonic decrease in the residual. In contrast, the second scenario displays much stronger nonlinearity due to the high Reynolds number, causing the residual to oscillate while still trending downward. The third scenario exhibits a mixed convergence pattern under different flow conditions. The neural operator is trained with just a few flow cases in all scenarios and the predictions on testing flows are used for initializing the steady-state simulations. The results demonstrate the efficacy of our approach, achieving acceleration ratios of at least two-fold while preserving the same level of accuracy. On the other hand, if we take the prediction of the trained neural operator as the solution to the PDEs directly, although the pointwise error is relatively small on average, the integral quantities (often of much more interest in engineering, such as drag and lift coefficients) may deviate more significantly from the values obtained from an accurate solution.

\subsection{Two-dimensional incompressible laminar flow}
\label{sec:laminar}

\begin{figure}[!htb]
\centering
\includegraphics[width=0.88\textwidth]{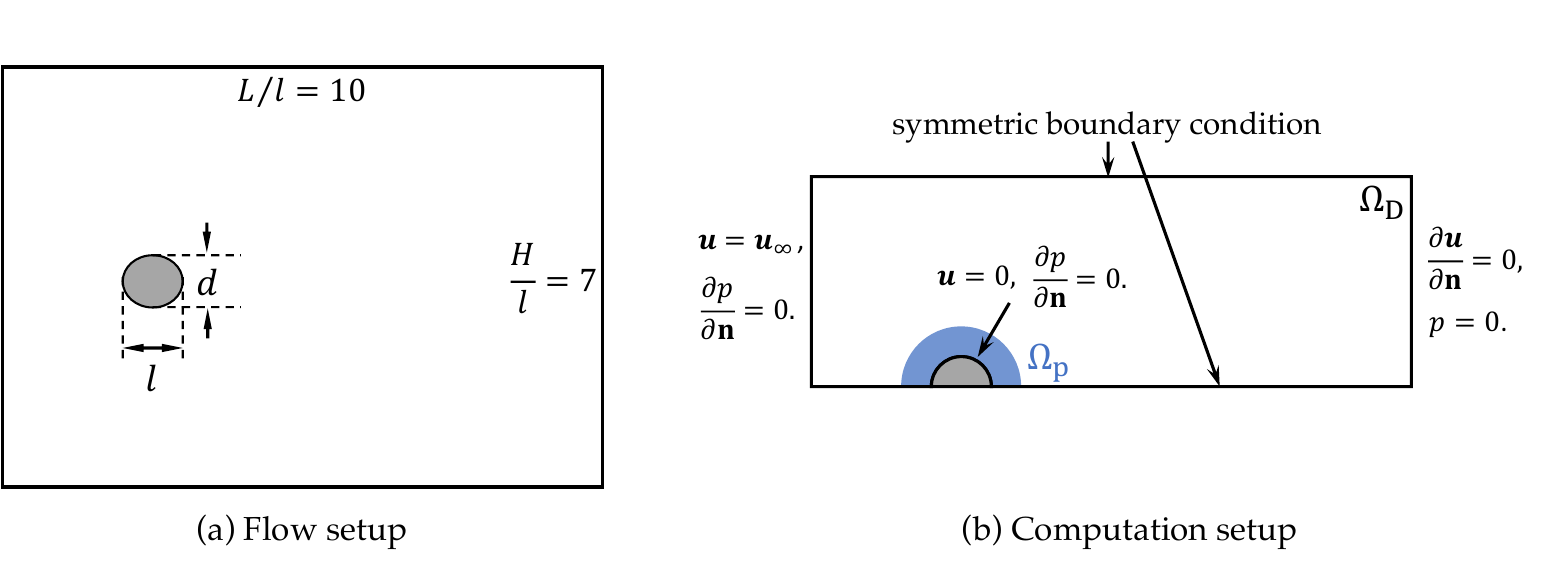}
  \caption{
  Setup for the incompressible laminar flows over parameterized elliptical cylinders: (a) flow setup, where the ratio of the thickness $d$ to the fixed length $l$ of the elliptical cylinder is adjusted between 0.75 and 1.3 to generate a family of laminar flows at different Reynolds numbers, and (b) computation setup showing the actual computational domain $\Omega_\text{D}$ with the boundary conditions. Here $\Omega_\text{p}$ refers to a region near the elliptical cylinder, where the distance from any point within this region to the cylinder wall is less than half of the cylinder's length $l$.
  }
  \label{fig:case1-setup}
\end{figure}

Our first scenario considers incompressible laminar flows over elliptical cylinders at different Reynolds numbers, which is illustrated in Fig.~\ref{fig:case1-setup}(a). The Reynolds number is defined as $Re = \mathsf{u}_{\infty} d/\nu$, where $\mathsf{u}_{\infty}$ and $\nu$ denote the freestream velocity magnitude and kinematic viscosity, respectively. We adjust the non-dimensionalized cylinder thickness, denoted as $d/l$, within the range of 0.75 to 1.3 to control the Reynolds number, maintaining it between 30 and 52. Specifically, we generate 12 flows with uniformly distributed Reynolds numbers: six training flows with $Re = [30, 34, \ldots, 50]$, five interpolated testing flows with $Re = [32, 36, \ldots, 48]$, and one extrapolated testing flow with $Re = 52$. 
Here, by assuming $\rho$ to be constant and the Stokes's stress constitutive equation $\bm{\tau}=\nu(\nabla \bm{u}+\nabla \bm{u}^\top)$, the governing equations~\eqref{eq:high-fidelity} and~\eqref{eq:NS-Eq} are reduced to the following steady-state N--S equations:
\begin{equation}
\begin{aligned}
\nabla \cdot \bu & =0, \\
\bu \cdot \nabla \bu-\nu \nabla^2 \bu+\frac{1}{\rho} \nabla p & =0,
\end{aligned}
\label{eq:NS-incompressible}
\end{equation}
where velocity $\bm{u}$ and pressure $p$ are to be solved for.
Therefore, the objective of super-fidelity is to learn the mapping from potential flows to steady-state incompressible laminar flows, i.e., from the solution of Eq.~\eqref{eq:low-fidelity} to Eq.~\eqref{eq:NS-incompressible}. The numerical simulations of potential and laminar flows are performed with the open-source CFD platform OpenFOAM~\cite{opencfd21openfoam}, employing the solvers \texttt{potentialFoam} and \texttt{simpleFoam}, respectively. We conduct the simulations on half of the full domain by implementing symmetric boundary condition on the bottom boundary, as shown in Fig.~\ref{fig:case1-setup}(b). This setting eliminates any potential unsteadiness that may arise in full-domain simulations of laminar flows. 
In the simulation of laminar flows, Eq.~\eqref{eq:NS-incompressible} is iteratively solved until the residuals satisfy the convergence criterion, i.e., $\cO < 10^{-8}$. In this study, residuals are defined in an absolute manner, as detailed in~\ref{app:res-defi}.

During the training phase, we do not utilize the potential flow field across the entire computational domain $\Omega_\text{D}$ as input for training efficiency. Instead, the input $\cQ$ consists of two parts: (1) the potential flow solution within the near-cylinder region $\Omega_\text{p}$, which is represented by the feature vectors attached to about 400 points sampled in this region, and (2) the potential flow solution at a target point where the laminar flow solution is to be predicted.
We specifically select this small set of pre-computed low-fidelity solutions, rather than full-domain solutions, as it is more physically informative compared to remaining regions. This strategy ensures that the training process remains computationally efficient while maintaining its effectiveness.
The desired output is the velocity and pressure of the laminar flow at the target point. Since $\Omega_\text{D}$ contains approximately 7000 cells, we have about 7000 input-output data pairs for each training or testing flow.
The neural operator is trained to minimize the misfit, using  mean squared error (squared $L_2$ norm), between its predictions and the iteratively converged laminar flow fields. The training process is performed on an open-source machine learning framework PyTorch~\cite{paszke2019pytorch}, which takes about half an hour on a single V100 graphics processing unit (GPU). More information regarding the training settings can be found in~\ref{app:NN-arci}.

\begin{figure}[!htb]
\centering
\includegraphics[width=0.88\textwidth]{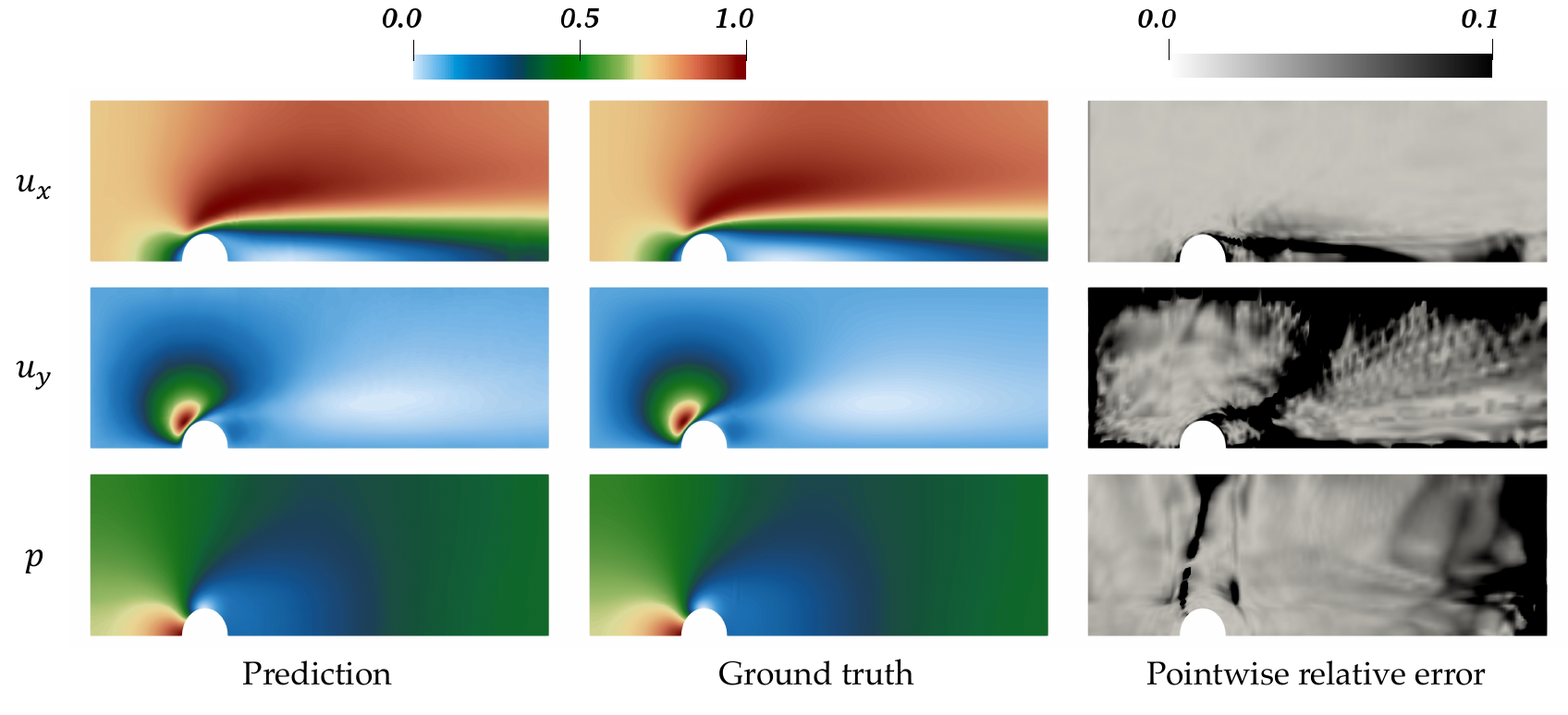}
  \caption{
  Comparison of the predicted solution (left column) and the corresponding ground truth (middle column), along with the plots of pointwise relative errors (right column) for the testing flow over an elliptical cylinder at $Re = 48$. The ground truth refers to the iteratively converged solution of velocity, $u_x$ and $u_y$, and pressure, $p$, at the residual of $10^{-8}$. The ground truths and predictions are non-dimensionalizaed and scaled between 0 and 1, while the pointwise relative errors are scaled between 0 and 10\% for clarity.
  }
  \label{fig:case1-compare}
\end{figure}

\begin{figure}[!htb]
\centering
\includegraphics[width=0.5\textwidth]{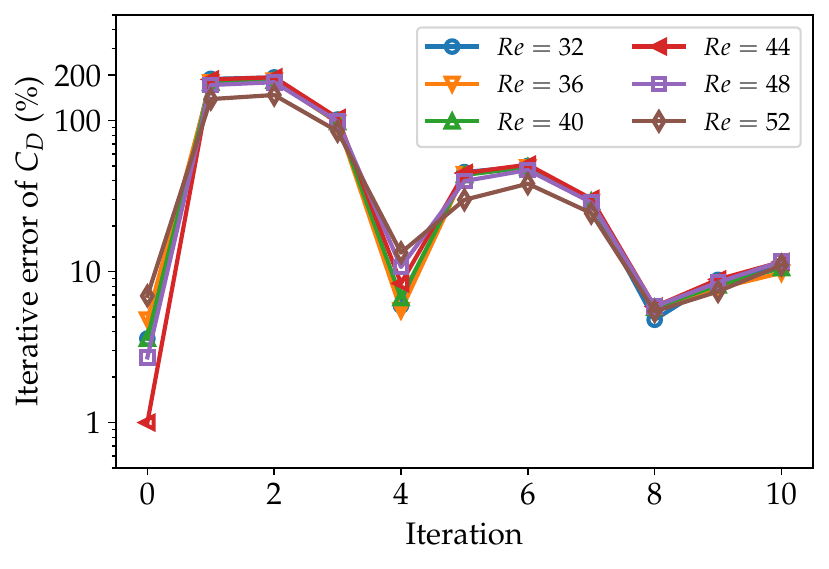}
  \caption{
  Iterative errors of drag coefficients in six testing flows over elliptical cylinders in the initial ten iterations after using neural network predictions as initial conditions.
  }
  \label{fig:case1-iter-err}
\end{figure}

The trained model exhibits good prediction performance on the testing flows, as shown in Fig.~\ref{fig:case1-compare}. The left column presents the predictions for the testing flow at $Re = 48$, which closely resemble the iteratively converged solutions displayed in the middle column. This prediction capability is further demonstrated by the small prediction errors in both solution and drag coefficient across all six testing flows, as provided in Table~\ref{Tab:case1-err}. In this work, the ground truths used to calculate prediction errors for solution and integral quantities refer to the solution achieved through iterative convergence at $\mathcal{R} = 10^{-8}$ and its corresponding integral quantities. However, it is important to note that while the model shows promise in prediction, the predicted solutions remain unreliable for direct application. Noticeable non-smoothness in the predictions is evident in the plots of pointwise relative error shown in the right column of Fig.~\ref{fig:case1-compare}. The relative errors, particularly those near the cylinder surface and other boundaries, may lead to violations of the boundary conditions. Moreover, the non-smoothness in predictions and non-compliance with boundary conditions becomes more apparent during the convergence process of drag coefficients when integrating the predictions into the solver for iterative computation, as shown in Fig.~\ref{fig:case1-iter-err}. Here, the iterative errors of drag coefficients can reach up to 200\% within the initial few iterative steps when using the predicted solutions as initial conditions. Consequently, traditional solvers are still necessary for achieving more accurate and reliable solutions.

\begin{table}[ht]
\centering
\caption{Prediction errors for six testing flows around elliptical cylinders, including five interpolated flows and one extrapolated flow. The relative $L_2$ norm errors for the solutions (velocity and pressure in $\Omega_\text{D}$) and the relative errors for the drag coefficients are presented.}
\begin{tabular}{ccccccc}
\toprule
\multicolumn{1}{c}{} & \multicolumn{5}{c}{\textbf{Interpolation}} & \multicolumn{1}{c}{\textbf{Extrapolation}} \\
\cmidrule(rl){2-6} \cmidrule(rl){7-7}
\textbf{Prediction error (\%)} & {$Re = 32$} & {$Re = 36$} & {$Re = 40$} & {$Re = 44$} & {$Re = 48$} & {$Re = 52$} \\
\midrule
Solution in $\Omega_\text{D}$ & 0.82 & 0.81 & 0.89 & 0.96 & 1.0 & 1.2\\
Drag coefficient, $C_\text{D}$ & 3.6  & 4.8 & 3.6 & 1.0 & 2.7 & 6.9 \\
\bottomrule
\end{tabular}
\label{Tab:case1-err}
\end{table}

\begin{figure}[!htb]
\centering
\includegraphics[width=0.99\textwidth]{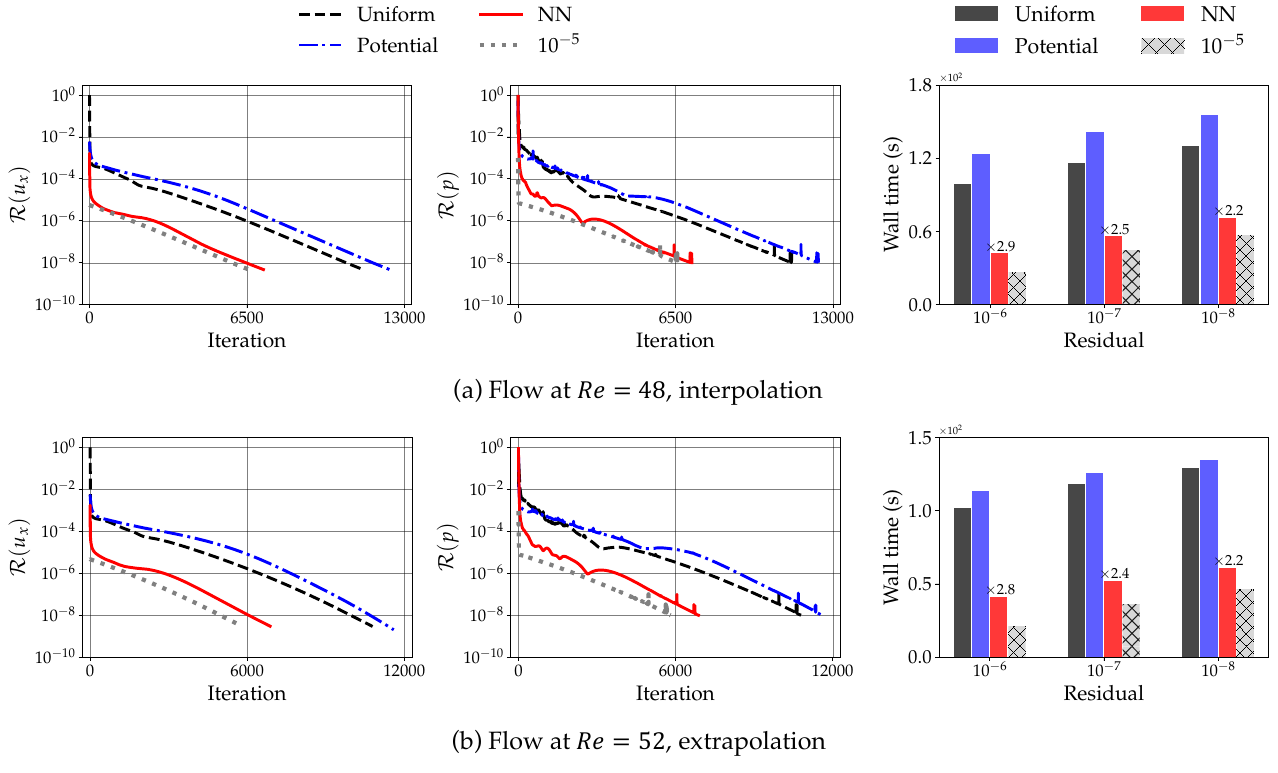}
  \caption{
  Comparison of three different initialization methods: uniform field, potential flow, and neural network prediction in both (a) interpolated and (b) extrapolated testing flows over elliptical cylinders. Left two columns: the convergence processes of velocity and pressure; right column: the wall time required to satisfy three convergence criteria ($10^{-6}$, $10^{-7}$, and $10^{-8}$). The legend $10^{-5}$ here refers to another refined initial condition at the residual of $10^{-5}$, offering a more comprehensive assessment of the acceleration performance of neural network predictions.
  }
  \label{fig:case1-residual}
\end{figure}

Using neural networks for initial conditions significantly accelerates iterative convergence, which is illustrated in Fig.~\ref{fig:case1-residual}. Specifically, we compare the iterative convergence processes using three different initial conditions: uniform field, potential flow, and neural network prediction. We conduct this comparison for both interpolated and extrapolated testing flows. It is evident from the first two columns of Fig.~\ref{fig:case1-residual} that using neural network predictions for initialization establishes an early lead in the convergence process in both testing cases and maintains this advantage consistently. As a result, it significantly reduces the number of iterations required to achieve convergence, thereby accelerating the iterative solution without loss of accuracy. For better showcasing this acceleration capability, we employ another refined initial condition for reference. Specifically, this refined initial condition is achieved by early stopping the simulation at a residual of $10^{-5}$ using the uniform field for initialization. We can observe that the convergence processes based on the neural network initialization and the refined initialization ($10^{-5}$) are quite close, with the former being slightly slower.
In other words, the neural network initialization in this scenario accelerates the solution process by effectively skipping the initial convergence phase, enabling the residual to approach approximately $10^{-5}$, with only marginal adjustments in the subsequent convergence rate.

This acceleration is more evidently reflected in the computational time, as is illustrated in the third column of Fig.~\ref{fig:case1-residual}. Specifically, we establish three different convergence criteria for residuals at $10^{-6}$, $10^{-7}$, and $10^{-8}$, and then compare the wall time associated with four different initial conditions. Note that the wall time covers the entire computational duration, including both pre-processing phase and iterative computation from initial state to final convergence. For potential flow initialization, the pre-processing time involves the time to solve Eq.~\eqref{eq:low-fidelity}; for neural network initialization, it involves solving Eq.~\eqref{eq:low-fidelity} and the additional time needed for neural network prediction; for initialization using solution at the residual of $10^{-5}$, we consider it an ideal initialization, using it merely as a reference, and assume there is no pre-processing time. 
In each testing flow, neural network initialization significantly reduces the wall time needed to achieve any of the convergence criteria.
To quantitatively evaluate this acceleration effect, we have introduced an acceleration ratio in this study, which is calculated as the wall time required when using potential flow initialization divided by that when using neural network initialization.
For the least demanding convergence criterion ($10^{-6}$), neural network initialization achieves an almost threefold acceleration, and for the strictest convergence criterion ($10^{-8}$), the acceleration is also more than twofold. While the ideal initialization shows further acceleration, it is usually achieved through a better trained neural network, which requires significantly larger training data volume and more training time. The comprehensive acceleration ratios for all testing flows over cylinders are listed in Table~\ref{tab:acceleration-s1} in~\ref{app:acceleration}.

\subsection{Two-dimensional turbulent flow}
\label{sec:turbulent}
Our second scenario considers two-dimensional turbulent flows over different airfoils at different angles of attack. The incoming Mach number remains constant at 0.15, and the Reynolds number based on chord length is $Re_\text{c} = 6 \times 10^6$. We select four airfoils (NACA0008, NACA0010, NACA0012, and NACA0015) and four angles of attack (9$^\circ$, 10$^\circ$, 12$^\circ$, and 13$^\circ$), denoted as $\alpha$, to generate 16 flows for training and testing. The testing flows are determined using Latin hypercube sampling~\cite{mckay2000comparison} to ensure the representativeness of testing flows, leading to 12 training flows and 4 testing flows. We also add the flow over NACA0011 at $\alpha=11^\circ$ for testing, where neither the airfoil nor the angle of attack are included in the training data. Overall, the five testing flows can be categorized into three types based on their relationship with the training flows: interpolation, weak interpolation, and extrapolation, as illustrated in Table.~\ref{tab:case2-design}. 
Here, by introducing a density-weighted time average decomposition of $\bm{u}$ and $e_t$, and a standard time average decomposition of $\rho$ and $p$, the governing equations~\eqref{eq:high-fidelity} and~\eqref{eq:NS-Eq} give the following equations, which is also known as the steady-state Farve-averaged Navier--Stokes equations~\cite{hirsch2007numerical}.
\begin{equation}
\begin{gathered}
\frac{\partial}{\partial x_j} (\bar{\rho} \hat{u}_j)=0, \\
\frac{\partial}{\partial x_j} (\hat{u}_j \bar{\rho} \hat{u}_i)=-\frac{\partial \bar{p}}{\partial x_i}+\frac{\partial \overline{\sigma}_{i j}}{\partial x_j}+\frac{\partial \tau_{i j}}{\partial x_j}, \\
\frac{\partial}{\partial x_j} [\hat{u}_j (\bar{\rho} \hat{e_t} + \bar{p})]=\frac{\partial}{\partial x_j}(\overline{\sigma}_{i j} \hat{u}_i+\overline{\sigma_{i j} u_i^{\prime \prime}})-\frac{\partial}{\partial x_j}\left(\bar{q}_j+c_p \overline{\rho u_j^{\prime \prime} T^{\prime \prime}}-\hat{u}_i \tau_{i j} + \frac{1}{2} \overline{\rho u_i^{\prime \prime} u_i^{\prime \prime} u_j^{\prime \prime}}\right).
\end{gathered}
\label{eq:farve-NS}
\end{equation}
The standard time (Reynolds) averages are denoted by an overbar, $\bar{\phi}$, while the density-weighted time (Favre) averages are denoted by a hat, $\hat{\phi}=\overline{\rho \phi} / \bar{\rho}$. Fluctuations around Reynolds and Favre averages are denoted by single and double primes, $\phi^{\prime}=\phi-\bar{\phi}$ and $\phi^{\prime \prime}=\phi-\hat{\phi}$, respectively. In Eq.~\eqref{eq:farve-NS}, the viscous stress $\bar{\sigma}_{ij}$, which corresponds to $\bm{\tau}$ in Eq.~\eqref{eq:NS-Eq}, is modeled as $\bar{\sigma}_{i j} = 2 \hat{\mu}(\hat{S}_{i j}-\frac{1}{3} \frac{\partial \hat{u}_k}{\partial x_k} \delta_{i j})$, while the Reynolds stress $\tau_{i j}$ ($\equiv-\overline{\rho u_i^{\prime \prime} u_j^{\prime \prime}}$), stemming from velocity fluctuations, is modeled as $\tau_{i j}=2 \hat{\mu}_t (\hat{S}_{i j}-\frac{1}{3} \frac{\partial \hat{u}_k}{\partial x_k} \delta_{i j})-\frac{2}{3} \bar{\rho} k \delta_{i j}$, where $\hat{S}_{i j}$ is the mean velocity strain, $k$ ($\equiv \frac{\widehat{u_i^{\prime \prime} u_i^{\prime \prime}}}{2}$) is the turbulent energy, $\hat{\mu}$ and $\hat{\mu}_t$ denote molecular and turbulent viscosity, respectively. Here, $\hat{\mu}_t$ is modeled through the Spalart-Allmaras turbulence model~\cite{spalart1992one}. The heat flux $\bar{q}_j$ and the turbulent heat flux $c_p \overline{\rho u_j^{\prime \prime} T^{\prime \prime}}$ are modeled as $-\frac{c_p \hat{\mu}}{P r} \frac{\partial \hat{T}}{\partial x_j}$ and $-\frac{c_p \hat{\mu}_t}{P r_t} \frac{\partial \hat{T}}{\partial x_j}$, respectively, where $c_p$ is the heat capacity at constant pressure, $\hat{T}$ is the temperature, $Pr$ and $Pr_t$ are laminar and turbulent Prandtl numbers, respective, with constant values of 0.72 and 0.85. The molecular diffusion term $\overline{\sigma_{i j} u_i^{\prime \prime}}$ and the turbulent transport term $-\frac{1}{2} \overline{\rho u_i^{\prime \prime} u_i^{\prime \prime} u_j^{\prime \prime}}$ are neglected considering their relatively small influences. To close Eq.~\eqref{eq:farve-NS}, the EOS is specified as
\begin{equation}
\bar{p} (\bar{\rho},\hat{e}) = (\gamma - 1) \bar{\rho} \hat{e}, \quad \text{with} \ \hat{e} = \hat{e_t} - \frac{1}{2} \hat{u}_i \hat{u}_i - k,
\label{eq:eos}
\end{equation}
where the heat capacity ratio $\gamma$ is taken as constant at 1.4, and $\hat{e}$ is related to $\hat{T}$ by $\hat{e} = \frac{1}{\gamma}c_p \hat{T}$. By incorporating Eq.~\eqref{eq:eos}, we solve Eq.~\eqref{eq:farve-NS} to obtain the velocity, pressure and internal energy per unit mass for the flows over airfoils.

\begin{table}[ht]
\centering
\caption{Training and testing cases in turbulent flows over airfoils. The check mark ($\bigcheck$) denotes a training flow while the circle mark ($\bm{O}$) represents a testing flow. The subscripts ``i'', ``wi'', and ``e'' refer to three testing types: interpolation, weak interpolation, and extrapolation.}
\label{tab:case2-design}
\renewcommand{\arraystretch}{0.8}
\begin{tabular}[t]{lC{2cm}C{2cm}C{2cm}C{2cm}C{2cm}}
\toprule
& NACA0008 & NACA0010 & NACA0011 & NACA0012 & NACA0015\\
\hline
$\alpha = 9^\circ$ & \bigcheck & \bigcheck & & $\bm{O}_\textbf{wi}$ & \bigcheck \\
$\alpha = 10^\circ$ & \bigcheck & \bigcheck & & \bigcheck & $\bm{O}_\textbf{wi}$ \\
$\alpha = 11^\circ$ & & & $\bm{O}_\textbf{i}$ & &  \\
$\alpha = 12^\circ$ & \bigcheck & $\bm{O}_\textbf{i}$ & & \bigcheck & \bigcheck \\
$\alpha = 13^\circ$ & $\bm{O}_\textbf{e}$ & \bigcheck & & \bigcheck & \bigcheck \\
\bottomrule
\end{tabular}
\end{table}

\begin{figure}[!htb]
\centering
\includegraphics[width=0.85\textwidth]{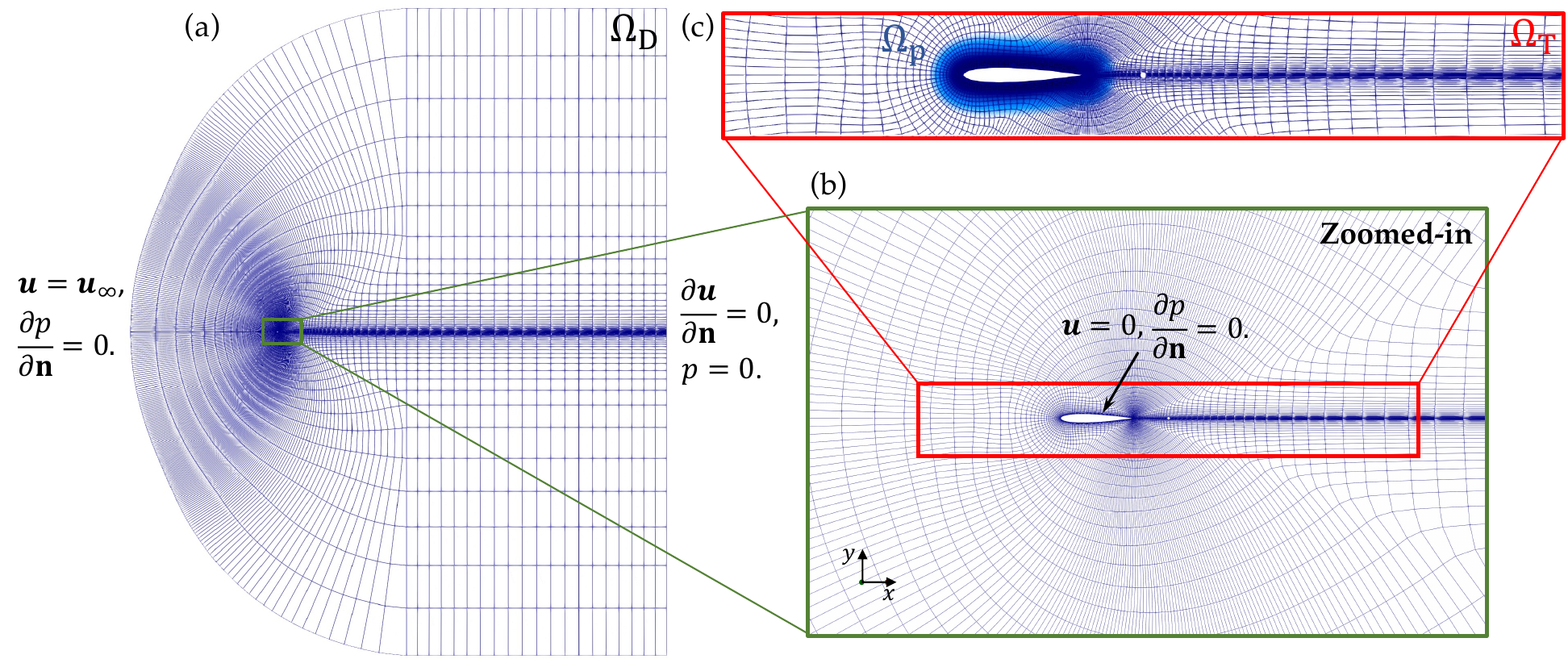}
  \caption{Computation setup for turbulent flows over airfoils: (a) computational domain $\Omega_\text{D}$ with boundary conditions for the inlet and outlet, (b) a zoomed-in region around airfoil with boundary conditions for the airfoil surface, and (c) training region $\Omega_\text{T}$, where steady-state turbulent flow fields are predicted, and a near-airfoil region $\Omega_\text{p}$ used for representing potential flows. The training region $\Omega_\text{D}$ extends horizontally to twice the chord length upstream and five times the chord length downstream, while vertically, it spans the length of one chord.
  The near-airfoil region $\Omega_\text{p}$ refers to a region where the distance from any point within this region to the airfoil surface is less than 0.3 times the chord length of the airfoil. Note that the height of the first mesh cell off the wall is sufficient small to ensure that the y-plus value remains below one.}
  \label{fig:case2-domain}
\end{figure}

The goal of super-fidelity here is to learn the mapping from potential flows to steady-state turbulent flows, i.e., from the solution of Eq.~\eqref{eq:low-fidelity} to Eq.~\eqref{eq:farve-NS}. The numerical simulations of turbulent flows are performed on a single process using the steady-state solver \texttt{rhoSimpleFoam} in OpenFOAM. The computational domain $\Omega_\text{D}$, discretized into approximately 32,000 cells, and the boundary conditions are illustrated in Fig.~\ref{fig:case2-domain}.
Note that despite the incoming Mach number being within the range for subsonic flows, we use the compressible solver in consideration of a potentially high local Mach number~\cite{nasa-airfoil}. For data generation, Eq.~\eqref{eq:farve-NS} is iteratively solved until the residuals satisfy the convergence criterion, i.e., $\mathcal{R} < 10^{-8}$.

During the training phase, we concentrate on predicting the turbulent flows in a specific region around the airfoils (see $\Omega_\text{T}$ in Fig.~\ref{fig:case2-domain}), while approximating the flows in the remaining region with freestream flows since the effect of airfoils out of $\Omega_\text{T}$ is assumed relatively small. Similar to the representation approach in the first scenario, the potential flow used as input does not encompass the entire domain. Specifically, the input includes two parts: (1) the potential flow solution within the near-airfoil region $\Omega_\text{p}$, which is represented by the feature vectors attached to about 4000 points sampled in this region, and (2) the potential flow solution at a target point where the RANS solution is to be predicted. The output consists of the velocity, pressure, and internal energy per unit mass at the target point. Since $\Omega_\text{T}$ consists of approximately 14000 cells, we have 14000 input-output data pairs for each training or testing flow. The training process on 12 flows is performed using PyTorch, which takes about 6.5 hours on a single V100 GPU. More information regarding the training settings is provided in~\ref{app:NN-arci}.

\begin{figure}[!htb]
\centering
\includegraphics[width=0.9\textwidth]{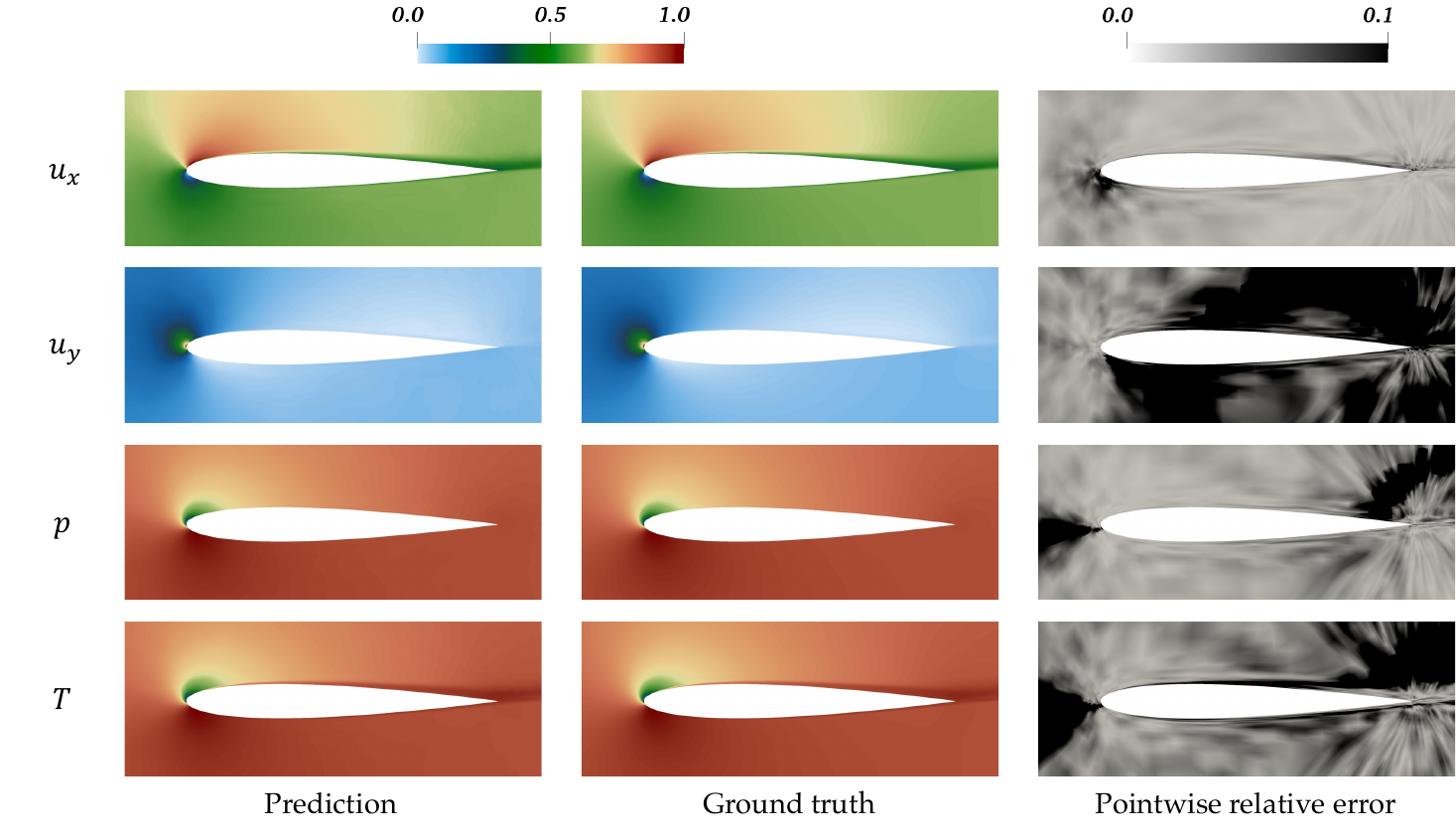}
  \caption{
  Comparison of the predicted solution (left column) and the corresponding ground truth (middle column), along with the plots of pointwise relative errors (right column) for the testing turbulent flow over NACA0011 at $\alpha = 11^\circ$. The ground truth refers to the iteratively converged solution at a residual of $10^{-8}$. The ground truth and prediction are non-dimensionalizaed and scaled between 0 and 1, while the pointwise relative errors are scaled between 0 and 10\% for clarity.
  }
  \label{fig:case2-compare}
\end{figure}

The trained model demonstrates good predictive performance even in this much more nonlinear scenario. In Fig.~\ref{fig:case2-compare}, we present a comparison between the neural operator predictions and the iteratively converged solutions for the testing flow over the NACA0011 airfoil at $\alpha = 11^\circ$. The predicted velocity, pressure, and temperature fields, which have been scaled between 0 and 1, exhibit a remarkable degree of visual similarity with their corresponding ground truth counterparts, as illustrated in the first two columns. This agreement is further demonstrated by the testing errors provided in Table~\ref{Tab:case2-err}. Notably, the prediction errors (i.e., relative $L_2$ norm errors) of solutions in $\Omega_\text{T}$ for all five testing flows remain consistently below 6\%. However, these predicted results have limited immediate applicability. Firstly, the presence of relatively high relative errors in specific regions, as shown in the third column of Fig.~\ref{fig:case2-compare}, can lead to non-smoothness and non-conservation for subsequent analyses. Secondly, while the prediction errors for the solutions are reasonably small, the integral quantities such as drag and lift coefficients may deviate more significantly from the iteratively converged values, as demonstrated in Table~\ref{Tab:case2-err}. In weak interpolation flows, despite the solution errors of approximately 3\%, the relative errors for drag coefficient exceed 10\%, which may not meet the precision standards in some application scenarios.

\begin{table}[!htp]
\centering
\caption{Prediction errors for five testing flows over airfoils, including two interpolated flows, two weakly interpolated flows, and one extrapolated flow. The relative $L_2$ norm errors for the solutions (velocity, pressure, and internal energy per unit mass in training region $\Omega_\text{T}$) and the relative errors for drag and lift coefficients are presented.}
\begin{tabular}{cccccc}
\toprule
\multicolumn{1}{c}{} & \multicolumn{2}{c}{\textbf{Interpolation}} & \multicolumn{2}{c}{\textbf{Weak interpolation}} & \multicolumn{1}{c}{\textbf{Extrapolation}} \\
\cmidrule(rl){2-3} \cmidrule(rl){4-5} \cmidrule(rl){6-6}
\textbf{Prediction error (\%)} & {NACA0010} & {NACA0011} & {NACA0012} & {NACA0015} & {NACA0008} \\
\midrule
Solution in $\Omega_\text{T}$ & 3.0 & 5.9 & 3.3 & 3.3 & 5.5  \\
Drag coefficient, $C_\text{D}$ & 3.6 & 8.6 & 15.9 & 11.4 & 6.6  \\
Lift coefficient, $C_\text{L}$ & 3.0 & 3.9 & 5.4 & 4.5 & 2.7  \\
\bottomrule
\end{tabular}
\label{Tab:case2-err}
\end{table}

\begin{figure}[!htb]
\centering
\includegraphics[width=0.99\textwidth]{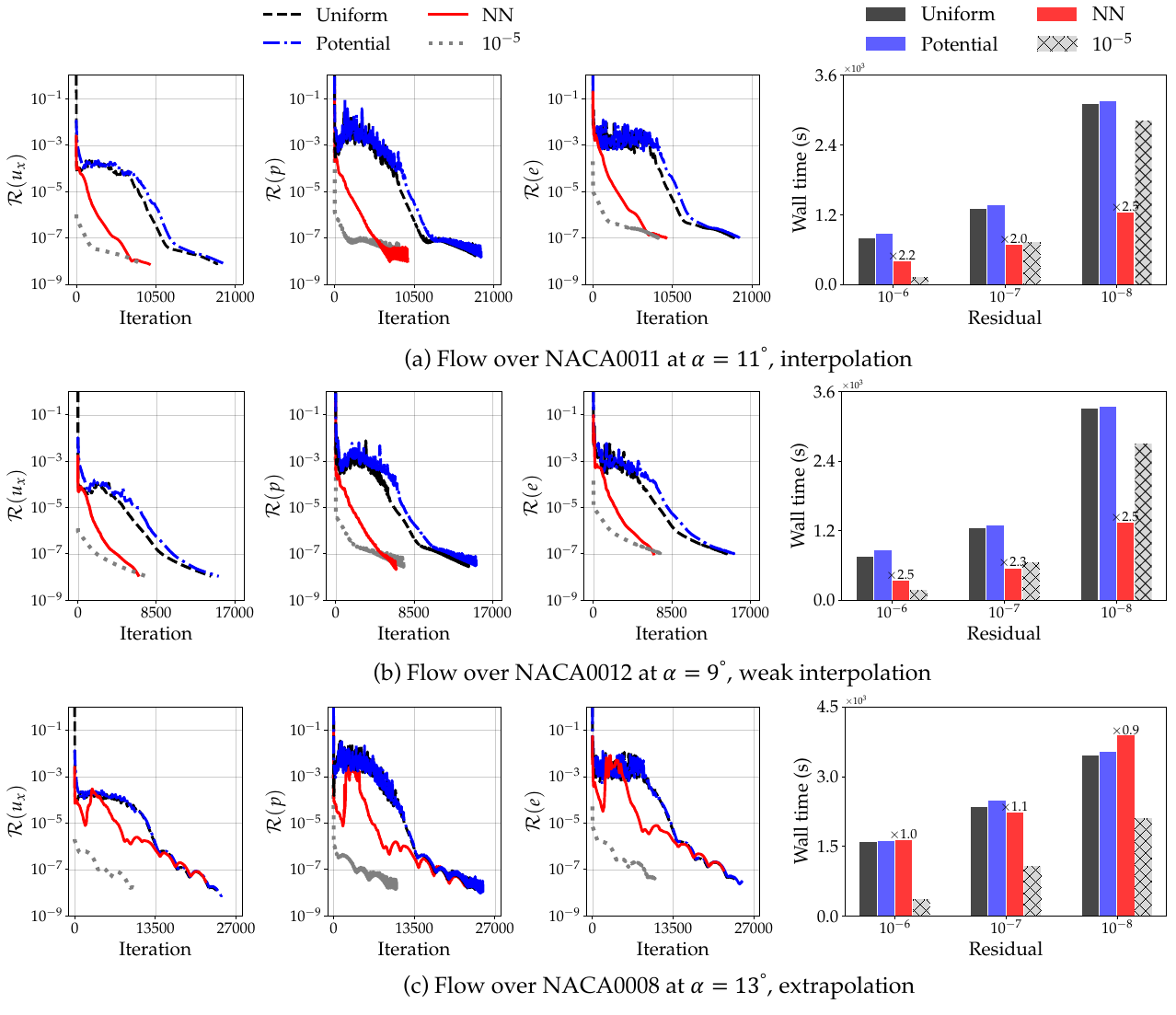}
  \caption{
  Comparison of three different initialization methods: uniform field, potential flow, and neural network prediction for (a) interpolated, (b) weakly interpolated, and (c) extrapolated testing flows over airfoils. Left three columns: the convergence processes of velocity, pressure, and internal energy per unit mass; right column: the wall time required to satisfy three convergence criteria ($10^{-6}$, $10^{-7}$, and $10^{-8}$). The legend $10^{-5}$ here refers to an ideal refined initial condition at the residual of $10^{-5}$, offering a more comprehensive assessment of the acceleration performance of neural network predictions.
  }
  \label{fig:case2-residual}
\end{figure}

To further enhance the solution precision, we employ the neural network predictions as initial conditions for solving Eq.~\eqref{eq:farve-NS} numerically. This approach exhibits a more robust convergence process and a notably faster convergence rate when compared to conventional initialization methods. In Fig.~\ref{fig:case2-residual}, we present a comparative analysis of convergence performance using three distinct initial conditions: uniform field, potential flow, and neural network prediction, across three different testing flow scenarios. In both interpolation and weak interpolation scenarios, it is evident that simulations initialized with neural network predictions require significantly fewer iterations to reach a specific residual level when compared to those initialized with uniform fields or potential flows.

The accelerated convergence can be attributed to two key factors. First, neural network initialization stabilizes early iterative convergence oscillations, as clearly seen in the iterative convergence processes shown in the left three columns of Fig.~\ref{fig:case2-residual}. Simulations initialized with neural network predictions effectively bypass the initial oscillatory phases that are present in simulations initialized with uniform fields or potential flows, transforming oscillatory convergence into a more stable and monotonic pattern. Second, neural network initialization improves average convergence rate. To illustrate this point, we introduce a refined initial condition, which is obtained by early stopping the numerical simulation at the residual of $10^{-5}$ using the uniform field for initialization. It is evident that the corresponding convergence processes also exhibit monotonic behavior without oscillations. However, the neural network initialization outperforms the refined initialization when aiming to achieve a relatively small convergence criterion (e.g., $10^{-8}$), as demonstrated by their convergence behaviors and the intersections of convergence curves.

The superiority of neural network initialization is more pronounced in the fourth column of Fig.~\ref{fig:case2-residual}, where the wall time for numerical simulations using different initialization methods is compared. For all three selected convergence criteria, the neural network initialization consistently achieves more than twofold acceleration ratios. Notably, the neural network initialization outperforms the refined initialization ($10^{-5}$) significantly for the convergence criterion $\mathcal{R} < 10^{-8}$, and slightly for $\mathcal{R} < 10^{-7}$. This advantage clearly illustrates that in highly nonlinear simulations, neural network initialization for acceleration does not depend solely on providing initial conditions with small residuals but also faster convergence during iterations, which is distinct from scenarios with weak nonlinearity where acceleration primarily depends on achieving a smaller initial residual. 

The neural network initialization approach does not present clear advantages over alternative initialization methods for the extrapolated testing flow, as illustrated in Figure~\ref{fig:case2-residual}(c). Despite an initial, brief lead due to the neural network prediction with a smaller residual, the subsequent sharp increase in residual significantly hinders the convergence. The abrupt increase in residual is likely attributable to changes in flow physics, particularly due to the stronger flow separation. Consequently, the wall time required for the simulation using neural network initialization is nearly identical to that of simulations employing uniform field or potential flow for initialization. Nevertheless, we still ensure the accuracy of solutions in extrapolated cases thanks to the engagement of CFD solvers. In contrast, achieving high accuracy in the extrapolation regime is a formidable task for neural network-based surrogate models, given their inherent nature of interpolation. The comprehensive acceleration ratios for all testing flows over airfoils are listed in Table~\ref{tab:acceleration-s2} in~\ref{app:acceleration}.

\begin{figure}[!htb]
\centering
\includegraphics[width=0.99\textwidth]{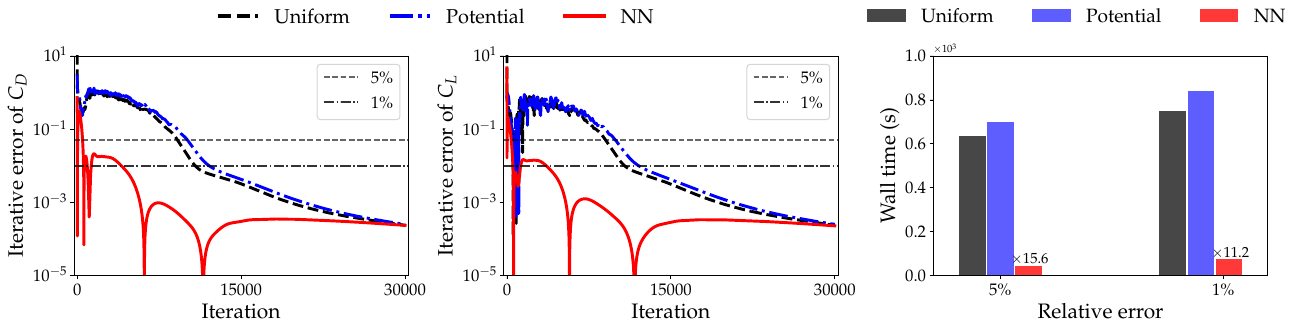}
  \caption{
  Comparison of three different initialization methods: uniform field, potential flow, and neural network prediction in the convergence of integral quantities. Left two panels: iterative errors for drag and lift coefficients; right panel: the wall time required to reach two relative error levels (5\% and 1\%).
  }
  \label{fig:coeffi-converge}
\end{figure}

Neural network initialization also significantly accelerates the convergence of integral quantities, as illustrated in Fig.~\ref{fig:coeffi-converge}. In the left and middle panels, we present the iterative errors of drag and lift coefficients for the interpolated testing flow over NACA0011 using three different initialization methods. Here, we consider the drag and lift coefficients at the convergence point, where the iterative residuals satisfy $\mathcal{R} < 10^{-8}$, as their ground truth values for calculating the iterative errors.
We can observe that, despite initial oscillations, the iterative errors of drag and lift coefficients achieved with neural network initialization rapidly converge to the 5\% and 1\% levels, exhibiting clear advantages over those resulting from the use of uniform fields or potential flows for initialization. Consequently, neural network initialization attains a 16-fold acceleration ratio for the 5\% error level and an 11-fold acceleration ratio for the 1\% error level for both drag and lift coefficients, as shown in the right panel.
It is important to note that monitoring iterative errors of integral quantities alone, rather than iterative residuals, can be misleading and should be employed cautiously in the numerical solution of PDEs. In this study, we investigate the acceleration performance in the convergence of integral quantities under the assumption of a more experienced scenario, where researchers/engineers have a high degree of confidence in the favorable behavior of iterative residuals.

\subsection{Three-dimensional turbulent flow}
We further evaluate the acceleration capability of the warm-start approach in a more complex and practical scenario by simulating three-dimensional turbulent flows over the ONERA M6 wing. The simulations cover incoming Mach numbers ranging from 0.45 to 0.75 and angles of attack between 0$^\circ$ and 6$^\circ$. To provide improved initial conditions and enable acceleration across this specified range, 12 flow conditions are selected for training: Mach numbers are sampled at increments of 0.1, and angles of attack are sampled at intervals of 3$^\circ$, ensuring comprehensive coverage of the parameter space.

\begin{figure}[!htb]
\centering
\includegraphics[width=0.8\textwidth]{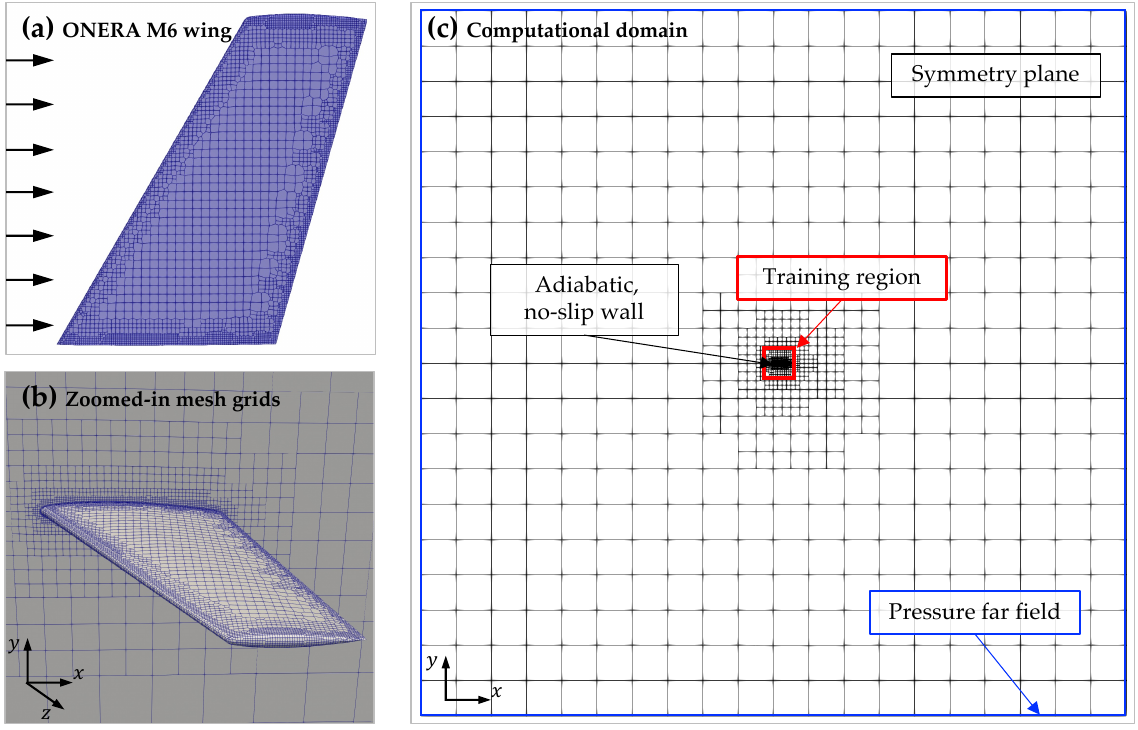}
  \caption{
  Computational setup for simulating turbulent flows over the ONERA M6 wing: (a) geometry of the ONERA M6 wing, (b) zoomed-in view of mesh near the wing, and (c) computational domain with boundary conditions. The highlighted box at the center of the domain represents the training region where RANS solutions are predicted.}
  \label{fig:case3-setup}
\end{figure}

The training process follows the steps described in the above two-dimensional turbulent flow case. Specifically, for the 12 selected flow conditions, potential flows and steady-state turbulent flows are simulated by solving the Laplace’s equation (Eq.~\eqref{eq:low-fidelity}) and the Farve-averaged Navier--Stokes equations (Eq.~\eqref{eq:farve-NS}) using the \texttt{potentialFoam} and \texttt{rhoSimpleFoam} solvers in OpenFOAM, respectively. The numerical simulations are performed on an unstructured mesh, as illustrated in Fig.~\ref{fig:case3-setup}. The ONERA M6 wing has a mean aerodynamic chord (MAC) length of 0.65 m and a span of 1.2 m. The computational domain extends 60 MAC lengths in the $x$- and $y$-directions and 30 MAC lengths in the $z$-direction and is discretized into 80,247 cells. While this mesh is relatively coarse, it provides reasonably accurate predictions for drag and lift~\cite{he2020dafoam}. The boundary conditions for turbulent flow simulations include an adiabatic no-slip wall on the wing surface, a symmetry condition at the inboard wing side, and a pressure far-field boundary located approximately 30 MAC lengths away. After simulating both flows, the neural operator is trained to map potential flow solutions to RANS solutions. Similarly, the training is confined to a near-wing region, as highlighted in Fig.~\ref{fig:case3-setup}(c), consisting of approximately 10,000 cell points, to efficiently capture critical turbulent flow features.

\begin{figure}[!htb]
\centering
\includegraphics[width=0.99\textwidth]{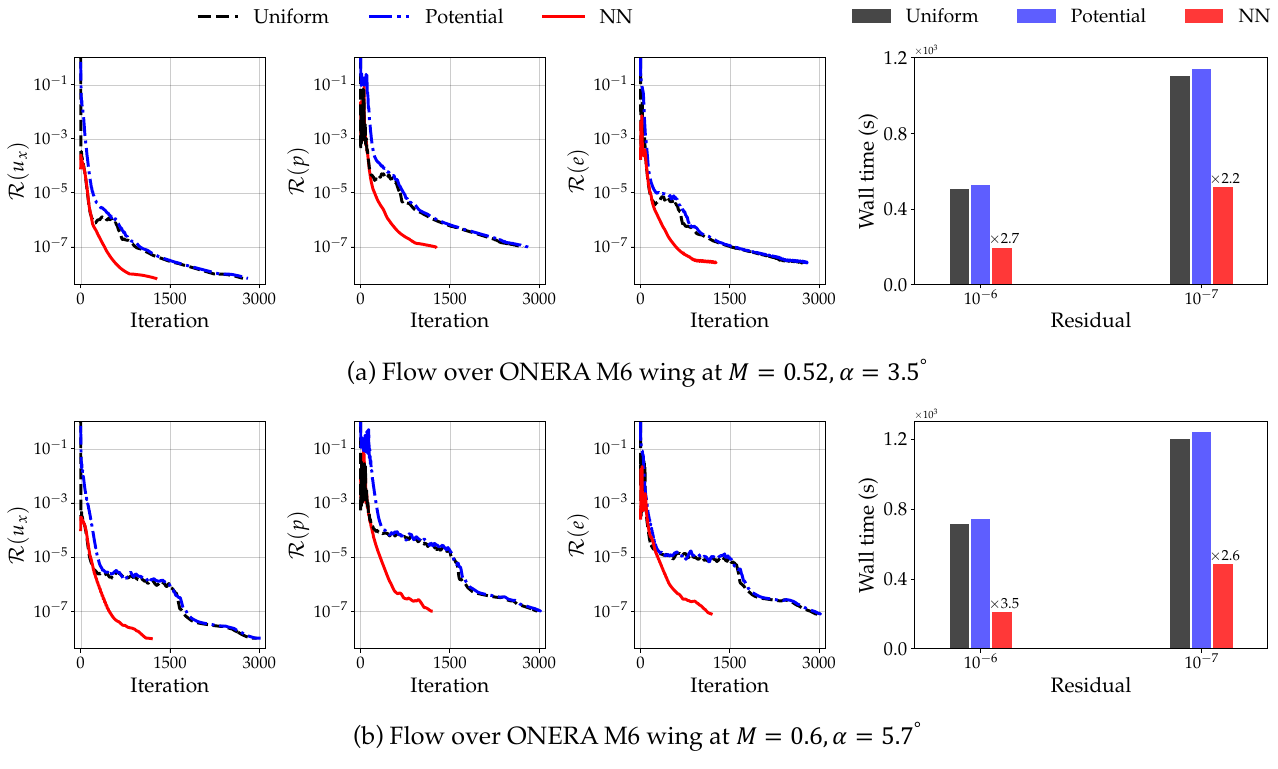}
  \caption{
  Comparison of three initialization methods: uniform field, potential flow, and neural operator prediction, for two testing flows over the ONERA M6 wing. Left three columns: convergence processes of velocity, pressure, and internal energy per unit mass. Right column: wall time required to reach residual thresholds ($10^{-6}$ and $10^{-7}$).
  }
  \label{fig:case3-residual}
\end{figure}

The trained neural operator effectively accelerates three-dimensional turbulent flow simulations, as illustrated in Fig.~\ref{fig:case3-residual}. Specifically, flow conditions are randomly selected from the parameter space for testing, and as examples, we present the convergence performance for two cases: (1) Mach number $M=0.52$ and angle of attack $\alpha=3.5^\circ$ and (2) Mach number $M=0.6$ and angle of attack $\alpha=5.7^\circ$.
Compared to the above two-dimensional turbulent flow case, which features a very fine mesh near the airfoil and exhibits prolonged initial-stage oscillations with uniform or potential flow initialization, this three-dimensional case achieves smoother and more monotonic convergence.
For the first test flow (top row), the convergence exhibits a steady decline throughout. In contrast, the second test flow (bottom row), characterized by a higher Mach number and angle of attack, experiences a noticeable slowdown during the intermediate stages. Nevertheless, the neural operator initialization consistently achieves significantly faster convergence for both test flows.
This is further illustrated in the last column, where the computational wall time required to reach two residual thresholds ($10^{-6}$ and $10^{-7}$) is compared across three initialization methods. The neural operator achieves consistent speedups exceeding two for both cases, with the acceleration becoming even more pronounced in the second case, surpassing three at the residual threshold of $10^{-6}$. These results highlight the neural operator’s efficiency in three-dimensional cases and its potential for accelerating practical flow simulations in industrial CFD applications.

\section{Discussion}
\label{sec:discuss}

\subsection{Acceleration for multiple simulations}

Despite the acceleration effect demonstrated in the above results, it is crucial to clarify that the warm-start model does not result in savings of total computation time for a single numerical simulation, as the time invested in data generation and model training far surpasses the time saved through improved initialization. Hence, we suggest employing this approach in scenarios involving multiple simulations to realize a substantial reduction in overall time. The reduction in total time depends primarily on two key factors: first, the time required to develop a warm-start model, and second, the number of numerical simulations needed after a neural operator is trained for warm-start. As such, we explore strategies for reducing model preparation time and investigate the impact of the number of simulations on the ultimate acceleration performance.

\begin{figure}[!htb]
\centering
\includegraphics[width=0.99\textwidth]{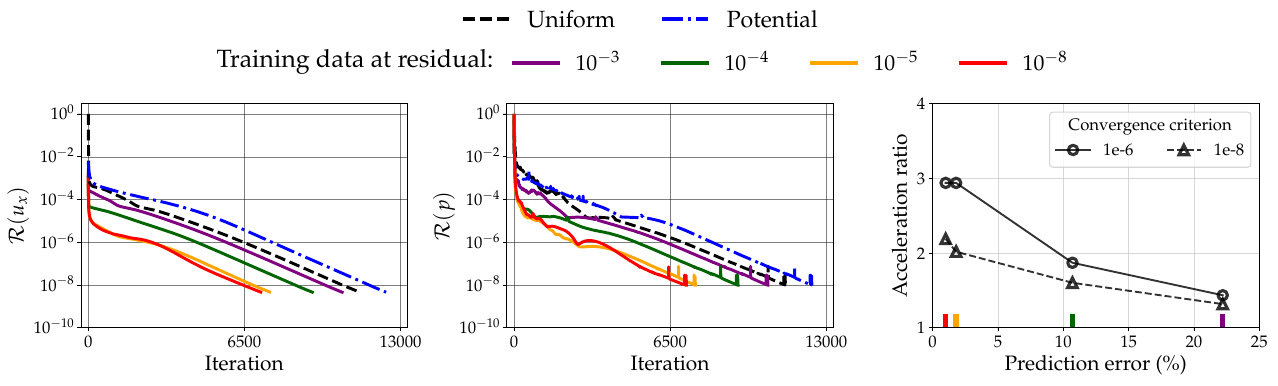}
  \caption{
  Comparison of the acceleration performance of the warm-start models trained with data of different quality in the testing flow over an elliptical cylinder. The left two panels show the corresponding convergence processes of velocity and pressure, respectively, while the right panel illustrates the relative $L_2$ norm errors of the predicted solutions associated with the four warm-start models (without additional CFD iterations) and the resulting acceleration ratios to reach two distinct convergence criteria.
  }
  \label{fig:case1-resi-diff-data}
\end{figure}

When preparing the warm-start model, we found that training the model with lower-quality data can lead to acceleration results comparable to those presented above. The significant advantage here lies in the considerably reduced time required for generating such lower-quality data. To investigate the influence of training data quality, we employed another three lower-quality datasets to train the neural operator and assessed their impact on acceleration performance in the aforementioned two flow scenarios. In Fig.~\ref{fig:case1-resi-diff-data}, we present the comparison in the testing flow around an elliptical cylinder. The warm-start models trained with data at residuals of $10^{-3}$, $10^{-4}$, and $10^{-5}$ exhibit increasing acceleration effects, with the last achieving results close to those from training with data at the residual of $10^{-8}$, as shown in the left two panels. In the third panel, we illustrate the performance of the corresponding four models, highlighting minimal prediction errors ($< 2\%$) and the similar, large acceleration ratios for the two warm-start models trained with data at residuals of $10^{-5}$ and $10^{-8}$, respectively.

\begin{figure}[!htb]
\centering
\includegraphics[width=0.99\textwidth]{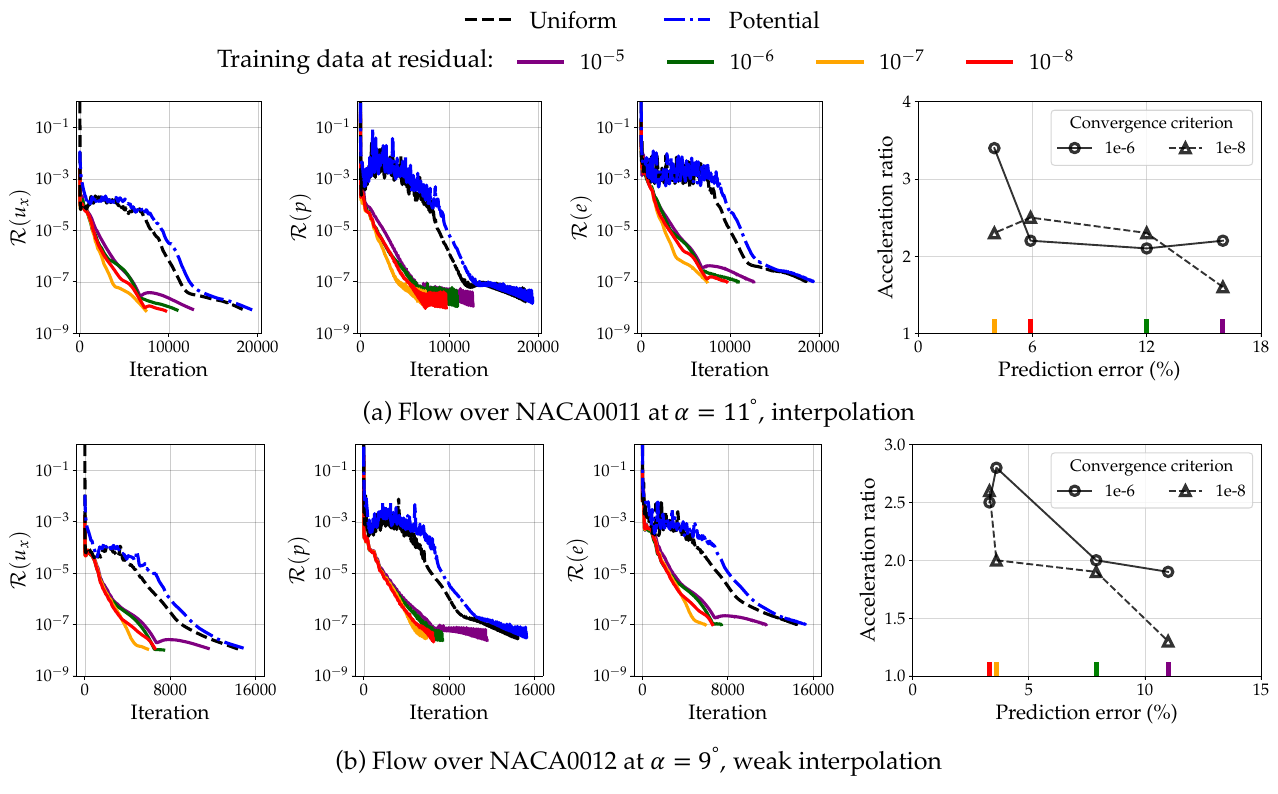}
  \caption{
  Comparison of the acceleration performance of the warm-start models trained with data of different quality in the testing flows over airfoils. The left three panels show the corresponding convergence processes of velocity, pressure, and internal energy per unit mass, respectively, while the right panel illustrates the relative $L_2$ norm errors of the predicted solutions associated with the four warm-start models (without additional CFD iterations) and the resulting acceleration ratios to reach two distinct convergence criteria.
  }
  \label{fig:case2-resi-diff-data}
\end{figure}

Similar effects of training data quality are also found in the second scenario for turbulent flows over airfoils, as shown in Fig.~\ref{fig:case2-resi-diff-data}. From the convergence processes illustrated in the first three columns, we can observe that all warm-start models trained with lower-quality data exhibit noticeable acceleration compared to those initialized with uniform or potential flows. Notably, the models trained with data at residuals of $10^{-6}$ and $10^{-7}$ display convergence processes that are similar to those based on the training data at the residual of $10^{-8}$. This similarity is more clearly shown in the fourth column. The relative $L_2$ norm error of predicted solution provided by the model trained with the data at the residual of $10^{-7}$ closely approximates that associated with the training data at the residual of $10^{-8}$, with both falling below 6\% in the two testing flows. Consequently, this similarity in prediction errors results in comparable acceleration ratios, both exceeding twofold improvement. On the other hand, while the prediction errors based on the model trained with the data at the residual of $10^{-6}$ nearly double, the corresponding acceleration ratios remain relatively stable, with the values about two. The results presented above demonstrate the practicality of training warm-start models using lower-quality data. This strategy reduces the time required for model development while preserving acceleration ratios of at least twofold.

We now present a potential application scenario for the warm-start approach. Based on the findings mentioned above, we assume that the warm-start model, developed to accelerate flow simulations over airfoils, is trained using a lower-quality dataset with a residual tolerance of $10^{-6}$ and is capable of achieving a twofold acceleration ratio. The total preparation time for the model is approximately 10 hours, comprising roughly 3.5 hours for data generation across 12 training flows and approximately 6.5 hours for neural network training. We have conducted a comparative analysis of the total time required for simulating different quantities of flows within the pipeline, using uniform fields for initialization versus neural network predictions for initialization. The results of this comparison are presented in Table~\ref{Tab:application}, where it is assumed that all simulations are required to converge to a residual threshold of $10^{-8}$.
It is evident that for a single numerical simulation ($N=1$), the time corresponding to neural network initialization is nearly ten times that based on uniform initialization due to the model preparation phase. As the number of simulations increases, the time corresponding to neural network initialization gradually approaches that based on uniform initialization, and subsequently gains a clear advantage. Notably, when $N=15$, the time for both initialization methods becomes almost identical. When $N=100$ and $N=1000$, the warm-start method achieves a substantial reduction in total time, offering acceleration ratios of approximately twofold, with the latter exhibiting even more pronounced time savings. Consequently, developing a neural operator-based warm-start model is well-justified in these scenarios.

It is important to note that this analysis only represents a hypothetical case. The actual acceleration achievable in practical applications requires further exploration. Nonetheless, based on our findings, we believe this method could provide significant benefits in various contexts. For example, in compressor blade design~\cite{zhao2019numerical}, the process involves a high-dimensional design space with parameters such as blade curvature, profile, thickness, and angle of attack. Iterative evaluations across numerous design configurations demand efficient initialization to accelerate convergence in diverse flow scenarios. Similarly, in shape optimization using ensemble-based methods~\cite{zhang2024large}, iterative updates require running simulations for multiple shapes at each iteration to approximate the optimization gradient. This process, often involving hundreds of simulations, can achieve significant computational savings with the warm-start approach. These examples illustrate the potential of the neural operator-based warm-start method to improve efficiency in practical design and optimization workflows.

\begin{table}[ht]
\centering
\caption{Comparison of total time required for varying simulation quantities: uniform initialization vs. neural network initialization. The time for model preparation, including 3.5 hours for data generation and 6.5 hours for model training, is considered in the neural network approach.}
\begin{tabular}{cccccc}
\toprule
\multicolumn{1}{c}{} & \multicolumn{5}{c}{\textbf{Number of simulations}} \\
\cmidrule(rl){2-6}
\textbf{Wall time comparison} & {$N = 1$} & {$N = 10$} & {$N = 15$} & {$N= 100$} & {$N = 1000$}  \\
\midrule
Uniform initialization (h) & 1.2 & 12 & 18 & 120 & 1200 \\
Neural operator initialization (h) & 10.5  & 15.3 & 18 & 63 & 540 \\
Acceleration ratio in total time & 0.1  & 0.8 & 1.0 & 1.9 & 2.2 \\
\bottomrule
\end{tabular}
\label{Tab:application}
\end{table}

\subsection{Robustness and scalability of the proposed methodology}
The results presented above are obtained using a single process with fixed linear equation solvers. To further evaluate the robustness of the proposed warm-start method, we examine its performance with different linear equation solvers available in OpenFOAM and extend the analysis to parallel computations with varying process counts. As examples, we evaluate the acceleration performance for flow over NACA0011 airfoil at $\alpha=11^\circ$ and flow over NACA0012 airfoil at $\alpha=9^\circ$, with the convergence criterion defined as $\mathcal{R} < 10^{-8}$.

\begin{figure}[!htb]
\centering
\includegraphics[width=0.8\textwidth]{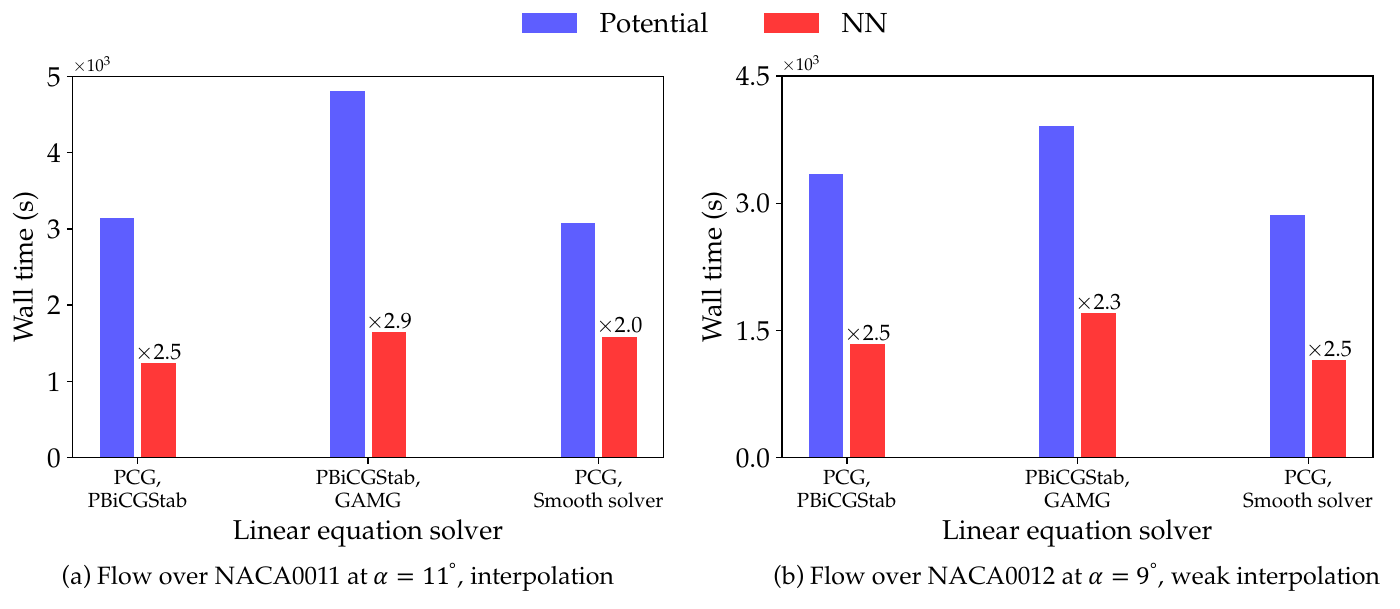}
  \caption{
  Acceleration performance across different linear equation solvers in interpolated and weakly interpolated testing flows over airfoils. Three configurations are compared: (1) PCG and PBiCGStab, (2) PBiCGStab and GAMG, and (3) PCG and smooth solver. In each configuration, the first solver is used for pressure, and the second is used for velocity and internal energy per unit mass.}
  \label{fig:case2-robust-solver}
\end{figure}

The warm-start method demonstrates its acceleration capabilities across various iterative algorithms, as illustrated in Fig.~\ref{fig:case2-robust-solver}. For the numerical solution following initialization, three different configurations of linear equation solvers are employed, with details provided in Table~\ref{Tab:diff-solver} in~\ref{app:diff-linear-solver}. The results demonstrate that neural network initialization consistently achieves a stable acceleration ratio, ranging from two to three, for each solver configuration in both interpolated and weakly interpolated test flows.

\begin{figure}[!htb]
\centering
\includegraphics[width=0.8\textwidth]{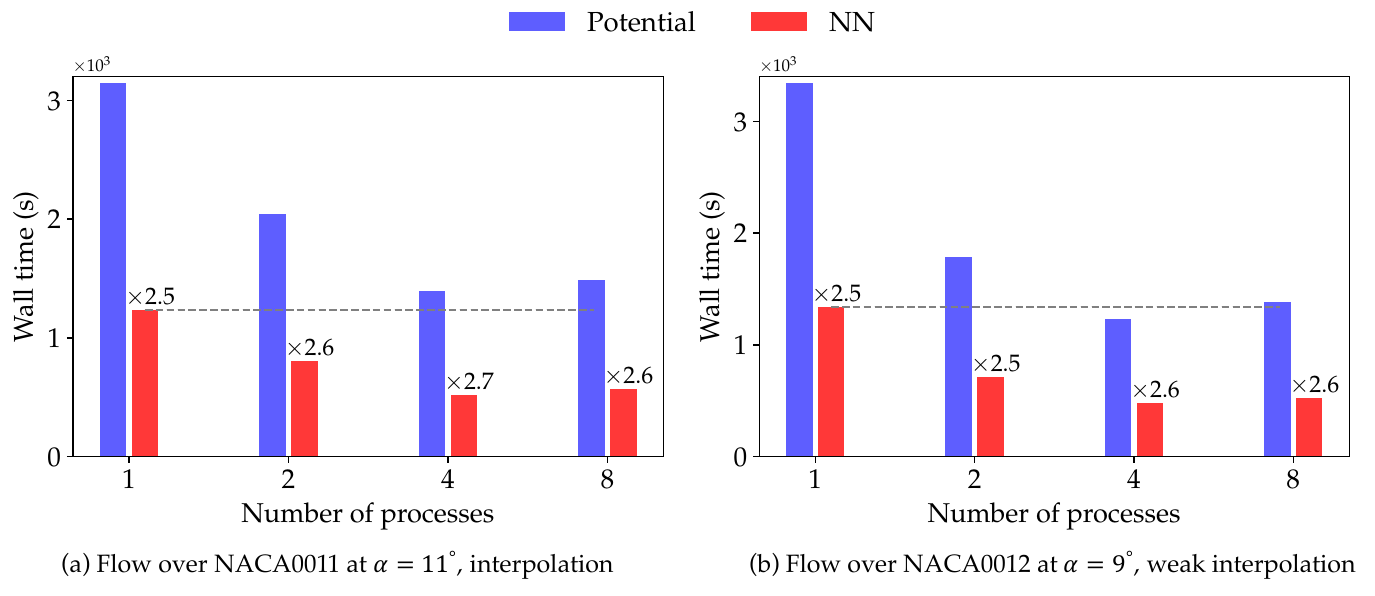}
  \caption{
  Acceleration performance across different process counts in interpolated and weakly interpolated testing flows over airfoils. The analysis covers a single process and parallel computing setups with two, four, and eight processes. Dashed lines highlight the performance of neural network initialization on a single process.}
  \label{fig:case2-robust-CPU}
\end{figure}

Similarly, the method delivers robust acceleration performance across parallel computing setups with different process counts, as shown in Fig.~\ref{fig:case2-robust-CPU}. The results indicate that neural network initialization consistently achieves an acceleration ratio of at least 2.5 compared to potential flow initialization. Notably, the convergence performance of neural network initialization on a single process, as highlighted by the dashed line, is comparable to that of potential flow initialization on four or eight processes, offering an alternative perspective on its acceleration capabilities. These findings collectively highlight the robustness and scalability of the warm-start method in reducing computational costs while maintaining consistent acceleration ratios across varying solver configurations and parallel computing setups.

While the acceleration ratios remain stable, the absolute gains decrease with increased parallelism, a trend that is expected to become more pronounced as hardware continues to evolve, including the potential adoption of GPU-accelerated solvers. However, it is important to note that the neural network training time can be significantly reduced by optimizing the training setup, such as increasing batch size and utilizing more powerful GPU platforms. These adjustments help maintain a stable overall acceleration ratio when considering training time.

\subsection{Comparison with other field-based initialization methods}
In the results presented above, we compare the neural operator-based initialization with uniform and potential flow initialization, two of the most commonly used approaches. Beyond these, other field-based prediction methods can also be considered for this purpose. This section examines several representative methods, including direct mapping, which uses the simulated flow field of the ``nearest'' case in the parameter space as the initial condition, as well as a discussion of the limitations of approaches based on convolutional neural networks (CNNs) and proper orthogonal decomposition (POD).

\begin{figure}[!htb]
\centering
\includegraphics[width=0.99\textwidth]{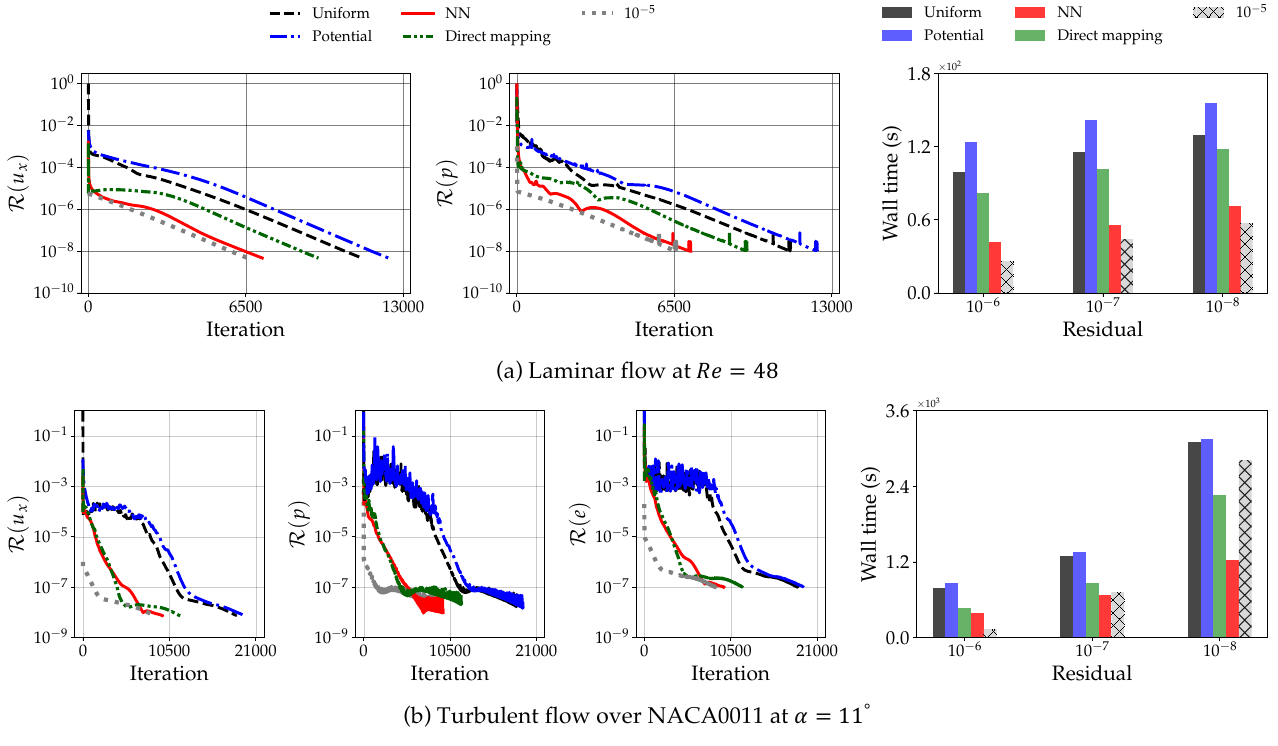}
  \caption{
  Comparison of direct mapping initialization with other methods for two test flows. Left two/three columns: iterative convergence processes; right column: computational wall time required to reach residual thresholds.
  }
  \label{fig:direct-mapping}
\end{figure}

First, we evaluate the acceleration achieved by the direct mapping method, as shown in Fig.~\ref{fig:direct-mapping}. This analysis focuses on two test flows: a laminar flow at $Re = 48$ and a turbulent flow over the NACA0011 airfoil at $\alpha = 11^\circ$. The direct mapping method initializes by interpolating the flow field from the nearest training case. For the laminar flow, the initial condition is derived from the flow field at $Re = 50$, while for the turbulent flow, it is obtained from the flow over the NACA0010 airfoil at $\alpha = 10^\circ$.
For the laminar flow (top row), the acceleration achieved by direct mapping is limited, offering only modest improvements compared to uniform and potential flow initializations. In contrast, the neural operator-based initialization achieves convergence in nearly half the time required by direct mapping, demonstrating significantly faster performance.
For the turbulent flow (bottom row), both direct mapping and neural operator-based initialization exhibit nearly identical convergence speeds during the early stages and substantially reduce the wall time needed to reach residual thresholds of $10^{-6}$ and $10^{-7}$ compared to uniform and potential flow initialization. However, for the more stringent threshold of $10^{-8}$, the neural operator shows a clear advantage, achieving faster convergence than direct mapping. 
Collectively, while direct mapping provides some acceleration compared to commonly used initialization methods and performs comparably to neural operator-based initialization in certain cases, the neural operator is overall superior in terms of convergence speed and robustness across flow conditions.

We further elaborate on the limitations of CNNs and POD to highlight the advantages of the neural operator-based initialization. While CNNs have been widely employed for learning field-to-field mappings~\cite{zhou2021learning,guo2016convolutional}, they face notable challenges in practical situations. First, CNNs are inherently designed for structured grids, rendering them unsuitable for unstructured meshes commonly encountered in industrial CFD applications. Second, CNNs are typically resolution-specific, requiring re-training or refining the model whenever spatial discretization of the input or output field changes, which significantly limits their flexibility and scalability. Similarly, regular POD-based methods are bound to the spatial resolution of the training snapshots, as the orthogonal basis functions are derived directly from the snapshot data. This dependency reduces their adaptability since any change in spatial discretization---such as switching to a finer or coarser grid---requires either re-computation of the POD modes or interpolation of the data onto a consistent grid~\cite{taira2017modal}. While these methods can provide effective initial conditions and achieve notable acceleration in specific scenarios, the neural operator-based approach offers a more robust and versatile alternative. By learning mappings directly in functional spaces, neural operators generalize across different grids and resolutions, enabling flexible and efficient initialization for diverse and complex flow cases.

\subsection{Limitation of neural networks/operators as surrogate models}
Directly using neural networks or neural operators as surrogate models has attracted increasing interest in various scientific simulations in recent years.
Despite the impressive predictive performance and fast evaluation speed, certain limitations of neural network-based surrogate models must be acknowledged. 
Specifically, the efficacy of neural network-based surrogate models thrives in scenarios characterized by a relatively lower dimensionality of the parameter space. When trained on extensive and diverse datasets/flows, these models demonstrate exceptional performance, enabling the discernment of intricate patterns and engendering accurate predictions. Nevertheless, the acquisition of such large-scale datasets demands substantial computational and data collection resources, which renders them impractical in applications with high-dimensional parameter spaces and limited computational capabilities. On the other hand, while these models perform well in fields covered by the training data, their efficacy diminishes considerably when extrapolating beyond the bounds of the training data. It is crucial to note that certain fields pose challenges in distinguishing between interpolation and extrapolation cases. Flows through porous media and urban canopies stand as prime examples~\cite{zhou2022neural,lu2023using}, where the heterogeneous composition of pores, voids, solid phases, and city-specific multi-scale land surface render the parameterization of geometries in these flows a difficult task. 

In contrast, our super-fidelity approach, which integrates classical solvers with neural operators, addresses these limitations. The classical iterative algorithm ensures accuracy and robustness in challenging scenarios, while the neural operator improves efficiency for cases well-represented in the training data.

\section{Conclusion}
\label{sec:conclude}

In this study, our primary objective is to accelerate the numerical simulation of a high-fidelity computational model described by steady-state PDEs without sacrificing solution accuracy. To achieve this, we introduce the idea of ``super-fidelity'', which refers to training of a neural operator to efficiently map solutions from a lower-fidelity model to the high-fidelity model. The lower-fidelity model allows for considerably faster computation, yet retains crucial physical insights. The solution predicted by the trained neural operator is not considered the final result; instead, it serves as an initial condition for solving the high-fidelity model, enhancing solving efficiency and ensuring result accuracy.

We have demonstrated the efficacy of this approach in three distinct scenarios: two-dimensional incompressible laminar flows around elliptical cylinders at low Reynolds numbers, two-dimensional turbulent flows over airfoils at a high Reynolds number, and practical three-dimensional turbulent flows over a wing. These scenarios represent typical scientific computing situations. In the first scenario characterized by weak nonlinearity, the iterative convergence exhibits a monotonic pattern, whereas in the second scenario marked by strong nonlinearity, it displays an oscillatory convergence pattern, and in the third scenario, it exhibits a mix of both behaviors. For these scenarios, we highlight the limitations of relying solely on neural operator predictions as final solutions, and assess the acceleration achieved by using neural operator predictions for initialization, comparing it with conventional initialization methods using uniform fields or potential flows.

The neural operator initialization delivers no less than a twofold acceleration while maintaining the same level of accuracy. It is noted that the acceleration behavior differs in these scenarios. In the first scenario with weak nonlinearity, the acceleration is attributed to the provision of initial condition with a smaller residual while the subsequent convergence rate remains nearly unchanged. In the second scenario marked by strong nonlinearity, the refined initial condition effectively transforms the oscillatory convergence into a more stable, monotonic pattern. Additionally, the average convergence rate also improves. Although the acceleration performance in extrapolated testing flows may not be as pronounced due to changes in flow physics, the accuracy of the results is still guaranteed, thanks to the iterations from traditional solvers. In the last practical scenario, the acceleration is attributed to a combination of these factors.

We suggest the application of this method in scenarios involving multiple simulations to achieve a substantial reduction in overall time, even when taking the time investment in model development into account. To further optimize the acceleration, we have introduced a practical strategy to reduce the time for model preparation. Our findings show that training the neural operator using lower-quality data still yields similar acceleration performance, while significantly reducing the time required for generating training data. 

\section*{Acknowledgments}
X.-H. Zhou is supported by the U.S. Air Force under agreement number FA865019-2-2204. The U.S. Government is authorised to reproduce and distribute reprints for Governmental purposes notwithstanding any copyright notation thereon. The computational resources used for this project were provided by the Advanced Research Computing (ARC) of Virginia Tech, which is gratefully acknowledged.

\appendix
\section{Neural operator architecture and training details}
\label{app:NN-arci}
Detailed neural operator (VCNN-e) architecture, consisting of two trainable sub-networks, is provided in Table~\ref{tab:nn-details}. The embedding network operates identically on the scalar features $\bm{c}^L$ attached to each point in the cloud $\cQ$ and outputs a row of $m$ elements in matrix~$\cG$. The fitting network maps the invariant feature matrix $\mathcal{D}$ to an invariant vector $\mathcal{E}$ and predictions of scalar quantities $\bm{c}^H$.

\begin{figure}[!htb]
\centering
\includegraphics[width=0.88\textwidth]{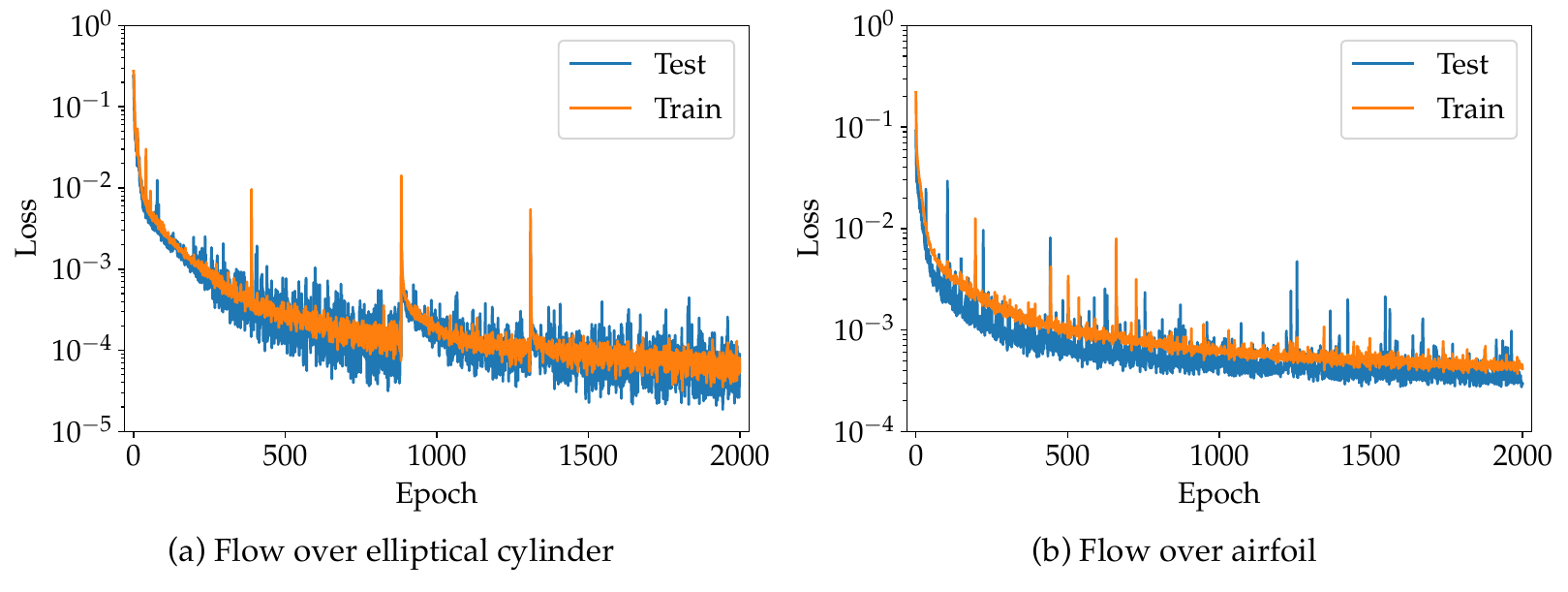}
  \caption{
  Training and testing losses for two different scenarios: (a) flows over elliptical cylinders at low Reynolds numbers, and (b) flows over airfoils at a high Reynolds number.
  }
  \label{fig:loss}
\end{figure}

The parameters of VCNN-e are optimized using the Adam optimizer~\cite{kingma2014adam}, with the standard mean squared error (MSE) as the loss function. In both flow scenarios, the training processes take 2000 epochs, with batch sizes of 256 and 128 for the first and second scenarios, respectively. The learning rate is set to be 0.001 at the beginning of the training process. The training and testing losses during the training processes are shown in Fig.~\ref{fig:loss}.

\begin{table}[htbp]
\caption{Detailed architectures of the embedding network and the fitting network in VCNN-e. The architecture specifies the number of neurons in each layer in sequence, from the input layer to the hidden layers (if any) and the output layer. The numbers of neurons in the input and output layers are highlighted in bold. Note that the number of output neurons in the fitting network for the second scenario (i.e., flows over airfoils) changes to 67 with two additional scalar features, internal energy per unit mass $e$ and turbulent eddy viscosity $\nu_t$, i.e., $\bm{c}^H \in \bbR^{3}$.}
\centering
\begin{tabular}{p{4.5cm} p{5cm} p{5cm}}
\toprule[1pt]
 & Embedding network ($\bm{c}^L \mapsto \mathcal{G}$) & Fitting network ($\mathcal{D} \mapsto \mathcal{E},\bm{c}^H$)\\
\midrule
No. of input neurons & 6 & $m\times m'= 256 \, (\cD \in \bbR^{64\times 4})$\\
No. of hidden layers & 3 & 2 \\
Architecture & 
(\textbf{6}, 16, 32, 64, \textbf{64}) 
& 
(\textbf{256}, 64, 64, \textbf{65})
\\
No. of output neurons & $m=64$ & 65 ($\mathcal{E} \in \bbR^{64}$, $\bm{c}^H \in \bbR$) \\
Activation functions & ReLU, Linear (last layer) & ReLU, Linear (last layer) \\
No. of trainable parameters 
& 6928 & 24833
\\
\bottomrule[1pt]
\end{tabular}
\label{tab:nn-details}
\end{table}

\section{Residual definition}
\label{app:res-defi}
In this study, we compare the iterative convergence of residual using different initialization methods. To ensure a fair and unbiased comparison, we calculate the residuals in an absolute manner, independent of the initial conditions. The residual calculation is as follows. For a matrix system
\begin{equation}
[\mathbf{A}][\boldsymbol{\Psi}]=[\mathbf{b}],
\end{equation}
the residual vector is defined as 
\begin{equation}
\mathbf{R} =\mathbf{b}-\mathbf{A} \cdot \boldsymbol{\Psi}.
\end{equation}
We apply residual scaling with the following normalisation procedure:
\begin{equation}
n_s = \|\mathbf{A} \cdot \mathbf{\Psi}-\mathbf{A} \cdot \overline{\mathbf{\Psi}}\|+\|\mathbf{b}-\mathbf{A} \cdot \overline{\mathbf{\Psi}}\|,
\end{equation}
where $\|\cdot\|$ is the matrix norm, calculated as the sum of the magnitude of each element therein, and $\overline{\mathbf{\Psi}}$ is the average of the solution vector. The scaled residual, i.e., the residual used throughout this paper, is finally given by
\begin{equation}
\begin{aligned}
\cO &= \frac{1}{n_s}\|\mathbf{R}\| \\
&= \frac{\|\mathbf{b}-\mathbf{A} \cdot \mathbf{\Psi}\|}{\|\mathbf{A} \cdot \mathbf{\Psi}-\mathbf{A} \cdot \overline{\bm{\Psi}}\|+\|\mathbf{b}-\mathbf{A} \cdot \overline{\bm{\Psi}}\|}.
\end{aligned}
\end{equation}
With this definition, the residual $\cO$ for a uniform field is always one, as $\bm{\Psi} = \overline{\bm{\Psi}}$ always holds true.

\section{Acceleration ratios across all testing flows}
\label{app:acceleration}
The comprehensive acceleration performance, achieved by neural operator initialization, is presented in Tables~\ref{tab:acceleration-s1} and~\ref{tab:acceleration-s2} for two distinct flow scenarios.

For testing flows around elliptical cylinders (see Table~\ref{tab:acceleration-s1}), the acceleration ratios consistently surpass twofold. They peak at nearly threefold when the residual tolerance is set at $10^{-6}$ and gradually decrease to twofold with a tolerance of $10^{-8}$. The decrease is in line with expectations, as the neural network prediction barely affects the convergence rate, resulting in a gradual erosion of the initial advantage when applying a stricter convergence criterion. 
For testing flows over airfoils (see Table~\ref{tab:acceleration-s2}), despite the increased complexity, the majority of acceleration ratios remain greater than twofold. However, due to the substantial nonlinearity therein, there is no discernible correlation between the acceleration ratios and the convergence criteria.

\begin{table}[ht]
\caption{Acceleration ratios for all six testing flows around elliptical cylinders using three different convergence criteria.}
\centering
\small
\begin{tabular}{ccccccc}
\toprule
\multicolumn{1}{c}{} & \multicolumn{5}{c}{\textbf{Interpolation}} & \multicolumn{1}{c}{\textbf{Extrapolation}} \\
\cmidrule(rl){2-6} \cmidrule(rl){7-7}
\textbf{Convergence criterion} & {$Re = 32$} & {$Re = 36$} & {$Re = 40$} & {$Re = 44$} & {$Re = 48$} & {$Re = 52$} \\
\midrule
$\cO(u_x), \cO(u_y), \cO(p) < 10^{-6}$ & 2.5 & 2.8 & 2.8 & 2.8 & 2.9 & 2.8 \\
$\cO(u_x), \cO(u_y), \cO(p) < 10^{-7}$ & 2.2 & 2.6 & 2.5 & 2.4 & 2.5 & 2.4 \\
$\cO(u_x), \cO(u_y), \cO(p) < 10^{-8}$ & 2.0 & 2.2 & 2.1 & 2.0 & 2.2 & 2.2 \\
\bottomrule
\end{tabular}
\label{tab:acceleration-s1}
\end{table}

\begin{table}[ht]
\caption{Acceleration ratios for all five testing flows over airfoils using three different convergence criteria.}
\centering
\small
\begin{tabular}{cccccc}
\toprule
\multicolumn{1}{c}{} & \multicolumn{2}{c}{\textbf{Interpolation}} & \multicolumn{2}{c}{\textbf{Weak interpolation}} & \multicolumn{1}{c}{\textbf{Extrapolation}} \\
\cmidrule(rl){2-3} \cmidrule(rl){4-5} \cmidrule(rl){6-6}
\textbf{Convergence criterion} & {NACA0010} & {NACA0011} & {NACA0012} & {NACA0015} & {NACA0008} \\
\midrule
$\cO(u_x), \cO(u_y), \cO(p), \cO(e) < 10^{-6}$ & 2.3 & 2.2 & 2.5 & 2.0 & 1.0  \\
$\cO(u_x), \cO(u_y), \cO(p), \cO(e) < 10^{-7}$ & 1.9 & 2.0 & 2.3 & 2.1 & 1.1  \\
$\cO(u_x), \cO(u_y), \cO(p), \cO(e) < 10^{-8}$ & 1.8 & 2.5 & 2.5 & 1.4 & 0.9  \\
\bottomrule
\end{tabular}
\label{tab:acceleration-s2}
\end{table}

\section{Linear equation solvers}
\label{app:diff-linear-solver}

To showcase the robustness of the warm-start approach, we have chosen three different configurations of linear equation solvers for the iterative solution after initialization. The linear equation solvers and associated preconditioner/smoothers in each configuration are presented in Table~\ref{Tab:diff-solver}.

In the first configuration, we use the preconditioned conjugate gradient (PCG) solver with the faster version of the diagonal-based incomplete Cholesky (FDIC) preconditioner for pressure and the preconditioned bi-conjugate gradient (PBiCGStab) solver with the diagonal-based incomplete LU (DILU) preconditioner for velocity and internal energy per unit mass. In the second configuration, we employ the PBiCGStab solver with the diagonal-based incomplete Cholesky (DIC) preconditioner for pressure and the geometric agglomerated algebraic multigrid (GAMG) solver with the symmetric Gauss Seidel (symGaussSeidel) smoother for velocity and internal energy per unit mass. In the third configuration, the PCG solver with the FDIC preconditioner is applied for pressure, while the smooth solver with the symGaussSeidel smoother is used for velocity and internal energy per unit mass. More details regarding the linear equation solvers, preconditioners, and smoothers can be found in OpenFOAM~\cite{opencfd21openfoam}.

\begin{table}[ht]
\centering
\caption{Three configurations of linear equation solvers used for demonstrating the robustness of the warm-start approach.}
\small
\begin{tabular}{ccccccc}
\toprule
\multicolumn{1}{c}{} & \multicolumn{2}{c}{\textbf{Configuration 1}} & \multicolumn{2}{c}{\textbf{Configuration 2}} & \multicolumn{2}{c}{\textbf{Configuration 3}} \\
\cmidrule(rl){2-3} \cmidrule(rl){4-5} \cmidrule(rl){6-7}
\textbf{} & {$p$} & {$\bm{u}\ |\ e$} & {$p$} & {$\bm{u}\ |\ e$} & {$p$} & {$\bm{u}\ |\ e$} \\
\midrule
Linear equation solver & PCG & PBiCGStab & PBiCGStab & GAMG & PCG & smoothSolver \\
Preconditioner & FDIC & DILU & DIC & - & FDIC & - \\
Smoother & - & - & - & symGaussSeidel & - & symGaussSeidel \\
\bottomrule
\end{tabular}
\label{Tab:diff-solver}
\end{table}


\begin{thebibliography}{10}
\expandafter\ifx\csname url\endcsname\relax
  \def\url#1{\texttt{#1}}\fi
\expandafter\ifx\csname urlprefix\endcsname\relax\def\urlprefix{URL }\fi
\expandafter\ifx\csname href\endcsname\relax
  \def\href#1#2{#2} \def\path#1{#1}\fi

\bibitem{han2019uniformly}
J.~Han, C.~Ma, Z.~Ma, W.~E, Uniformly accurate machine learning-based hydrodynamic models for kinetic equations, Proceedings of the National Academy of Sciences 116~(44) (2019) 21983--21991.

\bibitem{lu2021learning}
L.~Lu, P.~Jin, G.~Pang, Z.~Zhang, G.~E. Karniadakis, Learning nonlinear operators via {DeepONet} based on the universal approximation theorem of operators, Nature Machine Intelligence 3~(3) (2021) 218--229.

\bibitem{li2021fourier}
Z.~Li, N.~B. Kovachki, K.~Azizzadenesheli, B.~liu, K.~Bhattacharya, A.~Stuart, A.~Anandkumar, Fourier neural operator for parametric partial differential equations, in: International Conference on Learning Representations, 2021.

\bibitem{kovachki2023neural}
N.~B. Kovachki, Z.~Li, B.~Liu, K.~Azizzadenesheli, K.~Bhattacharya, A.~M. Stuart, A.~Anandkumar, Neural operator: Learning maps between function spaces with applications to {PDE}s., Journal of Machine Learning Research 24~(89) (2023) 1--97.

\bibitem{han2023equivariant}
J.~Han, X.-H. Zhou, H.~Xiao, An equivariant neural operator for developing nonlocal tensorial constitutive models, Journal of Computational Physics 488 (2023) 112243.

\bibitem{chen2023operator}
C.~Chen, J.-L. Wu, Operator learning for continuous spatial-temporal model with a hybrid optimization scheme, arXiv preprint arXiv:2311.11798 (2023).

\bibitem{huang2024operator}
D.~Z. Huang, N.~H. Nelsen, M.~Trautner, An operator learning perspective on parameter-to-observable maps, arXiv preprint arXiv:2402.06031 (2024).

\bibitem{zhou2024bi}
X.-H. Zhou, Z.-R. Liu, H.~Xiao, {BI-EqNO}: Generalized approximate {B}ayesian inference with an equivariant neural operator framework, arXiv preprint arXiv:2410.16420 (2024).

\bibitem{shukla2023deep}
K.~Shukla, V.~Oommen, A.~Peyvan, M.~Penwarden, L.~Bravo, A.~Ghoshal, R.~M. Kirby, G.~E. Karniadakis, Deep neural operators can serve as accurate surrogates for shape optimization: a case study for airfoils, arXiv preprint arXiv:2302.00807 (2023).

\bibitem{chen2023towards}
L.-W. Chen, N.~Thuerey, Towards high-accuracy deep learning inference of compressible flows over aerofoils, Computers \& Fluids 250 (2023) 105707.

\bibitem{raissi2019physics}
M.~Raissi, P.~Perdikaris, G.~E. Karniadakis, Physics-informed neural networks: A deep learning framework for solving forward and inverse problems involving nonlinear partial differential equations, Journal of Computational physics 378 (2019) 686--707.

\bibitem{grossmann2023can}
T.~G. Grossmann, U.~J. Komorowska, J.~Latz, C.-B. Sch{\"o}nlieb, Can physics-informed neural networks beat the finite element method?, arXiv preprint arXiv:2302.04107 (2023).

\bibitem{mcgreivy2024weak}
N.~McGreivy, A.~Hakim, Weak baselines and reporting biases lead to overoptimism in machine learning for fluid-related partial differential equations, Nature Machine Intelligence 6~(10) (2024) 1256--1269.

\bibitem{dong2015image}
C.~Dong, C.~C. Loy, K.~He, X.~Tang, Image super-resolution using deep convolutional networks, IEEE transactions on pattern analysis and machine intelligence 38~(2) (2015) 295--307.

\bibitem{kochkov2021machine}
D.~Kochkov, J.~A. Smith, A.~Alieva, Q.~Wang, M.~P. Brenner, S.~Hoyer, Machine learning--accelerated computational fluid dynamics, Proceedings of the National Academy of Sciences 118~(21) (2021) e2101784118.

\bibitem{um2020solver}
K.~Um, R.~Brand, Y.~R. Fei, P.~Holl, N.~Thuerey, Solver-in-the-loop: Learning from differentiable physics to interact with iterative {PDE}-solvers, Advances in Neural Information Processing Systems 33 (2020) 6111--6122.

\bibitem{zhou2022frame}
X.-H. Zhou, J.~Han, H.~Xiao, Frame-independent vector-cloud neural network for nonlocal constitutive modeling on arbitrary grids, Computer Methods in Applied Mechanics and Engineering 388 (2022) 114211.

\bibitem{li2019data}
J.~Li, M.~A. Bouhlel, J.~R. Martins, Data-based approach for fast airfoil analysis and optimization, AIAA Journal 57~(2) (2019) 581--596.

\bibitem{ma2023efficient}
W.~Ma, X.~Zhao, S.~Islam, A.~Narkhede, K.~Wang, Efficient solution of bimaterial {R}iemann problems for compressible multi-material flow simulations, Journal of Computational Physics 493 (2023) 112474.

\bibitem{hirsch2007numerical}
C.~Hirsch, Numerical computation of internal and external flows: The fundamentals of computational fluid dynamics, Elsevier, 2007.

\bibitem{ansys2013ansys}
{ANSYS Inc}, {ANSYS Fluent Theory Guide (Release 15.0)}, ANSYS Inc (2013).

\bibitem{opencfd21openfoam}
{The OpenFOAM Foundation}, \href{https://cfd.direct/openfoam/user-guide}{{OpenFOAM} User Guide} (2020).
\newline\urlprefix\url{https://cfd.direct/openfoam/user-guide}

\bibitem{zafar2022frame}
M.~I. Zafar, J.~Han, X.-H. Zhou, H.~Xiao, Frame invariance and scalability of neural operators for partial differential equations, Communications in Computational Physics 32~(2) (2022) 336--363.

\bibitem{oberkampf2010verification}
W.~L. Oberkampf, C.~J. Roy, Verification and validation in scientific computing, Cambridge university press, 2010.

\bibitem{phillips2014richardson}
T.~S. Phillips, C.~J. Roy, Richardson extrapolation-based discretization uncertainty estimation for computational fluid dynamics, Journal of Fluids Engineering 136~(12) (2014) 121401.

\bibitem{paszke2019pytorch}
A.~Paszke, S.~Gross, F.~Massa, A.~Lerer, J.~Bradbury, G.~Chanan, T.~Killeen, Z.~Lin, N.~Gimelshein, L.~Antiga, et~al., {PyTorch}: An imperative style, high-performance deep learning library, Advances in Neural Information Processing Systems 32 (2019).

\bibitem{mckay2000comparison}
M.~D. McKay, R.~J. Beckman, W.~J. Conover, A comparison of three methods for selecting values of input variables in the analysis of output from a computer code, Technometrics 42~(1) (2000) 55--61.

\bibitem{spalart1992one}
P.~Spalart, S.~Allmaras, A one-equation turbulence model for aerodynamic flows, in: 30th Aerospace Sciences Meeting and Exhibit, 1992, p. 439.

\bibitem{nasa-airfoil}
NASA, {2D NACA 0012 Airfoil Validation Case}, \url{https://turbmodels.larc.nasa.gov/naca0012_val.html}.

\bibitem{he2020dafoam}
P.~He, C.~A. Mader, J.~R. Martins, K.~J. Maki, Dafoam: An open-source adjoint framework for multidisciplinary design optimization with openfoam, AIAA journal 58~(3) (2020) 1304--1319.

\bibitem{zhao2019numerical}
X.~Zhao, X.~Zhou, J.~Cheng, J.~Chen, Numerical investigation on arbitrary polynomial blade model for a transonic axial-flow compressor rotor with multi-parameter optimization, in: The Proceedings of the 2018 Asia-Pacific International Symposium on Aerospace Technology (APISAT 2018) 9th, Springer, 2019, pp. 19--33.

\bibitem{zhang2024large}
X.-L. Zhang, F.~Zhang, Z.~Li, X.~Yang, G.~He, Large-eddy simulation-based shape optimization for mitigating turbulent wakes of a bluff body using the regularized ensemble {K}alman method, Journal of Fluid Mechanics 1001 (2024) A31.

\bibitem{zhou2021learning}
X.-H. Zhou, J.~Han, H.~Xiao, Learning nonlocal constitutive models with neural networks, Computer Methods in Applied Mechanics and Engineering 384 (2021) 113927.

\bibitem{guo2016convolutional}
X.~Guo, W.~Li, F.~Iorio, Convolutional neural networks for steady flow approximation, in: Proceedings of the 22nd ACM SIGKDD international conference on knowledge discovery and data mining, 2016, pp. 481--490.

\bibitem{taira2017modal}
K.~Taira, S.~L. Brunton, S.~T. Dawson, C.~W. Rowley, T.~Colonius, B.~J. McKeon, O.~T. Schmidt, S.~Gordeyev, V.~Theofilis, L.~S. Ukeiley, Modal analysis of fluid flows: An overview, Aiaa Journal 55~(12) (2017) 4013--4041.

\bibitem{zhou2022neural}
X.-H. Zhou, J.~E. McClure, C.~Chen, H.~Xiao, Neural network--based pore flow field prediction in porous media using super resolution, Physical Review Fluids 7~(7) (2022) 074302.

\bibitem{lu2023using}
Y.~Lu, X.-H. Zhou, H.~Xiao, Q.~Li, Using machine learning to predict urban canopy flows for land surface modeling, Geophysical Research Letters 50~(1) (2023) e2022GL102313.

\bibitem{kingma2014adam}
D.~P. Kingma, J.~Ba, Adam: A method for stochastic optimization, arXiv preprint arXiv:1412.6980 (2014).

\end{thebibliography}
\end{document}